\documentclass[manuscript, twocolumn,trackchanges,twocolappendix]{aastex701}
\usepackage{multirow}
\usepackage{makecell}
\usepackage{comment}
\usepackage{amsmath}
\usepackage{rotating}  
\usepackage{siunitx}   
\usepackage{booktabs}  
\usepackage{multirow}  

\begin{document}

\title{Science with a large field-of-view polarization survey: \\ The Large Array Survey Telescope Polarization Node (LAST-P)}

\author[orcid=0000-0002-5085-8828, gname=V., sname='Barbosa Martins']{V. Barbosa Martins}
\affiliation{Ruhr University Bochum, Faculty of Physics and Astronomy, Astronomical Institute (AIRUB), 44780 Bochum, Germany}
\affiliation{Instituto de Física, Universidade Federal de Goiás, 74690-900, Goiânia, GO, Brazil}
\email[show]{barbosamartins@ufg.br}

\author[orcid=0000-0002-5467-8277, gname='N. Jordana', sname=Mitjans]{N. Jordana Mitjans}
\affiliation{Ruhr University Bochum, Faculty of Physics and Astronomy, Astronomical Institute (AIRUB), 44780 Bochum, Germany}
\email{jordana@astro.ruhr-uni-bochum.de}

\author[orcid=0000-0003-2403-4582, gname=S., sname=Garrappa]{S. Garrappa}
\affiliation{Department of particle physics and astrophysics, Weizmann Institute of Science, 76100 Rehovot, Israel}
\email{simone.garrappa@gmail.com}

\author[orcid=0000-0002-5605-2219, gname=A., sname=Franckowiak]{A. Franckowiak}
\affiliation{Ruhr University Bochum, Faculty of Physics and Astronomy, Astronomical Institute (AIRUB), 44780 Bochum, Germany}
\email{franckowiak@astro.ruhr-uni-bochum.de}

\author[orcid=0000-0002-6786-8774, gname='E. O.', sname=Ofek]{E. O. Ofek}
\affiliation{Department of particle physics and astrophysics, Weizmann Institute of Science, 76100 Rehovot, Israel}
\email{eran.ofek@weizmann.ac.il}

\author[orcid=0000-0001-6760-3074, gname=S., sname=Ben-Ami]{S. Ben-Ami}
\affiliation{Department of particle physics and astrophysics, Weizmann Institute of Science, 76100 Rehovot, Israel}
\email{sagi.ben-ami@weizmann.ac.il}

\author[gname=J., sname='Borowska-Naguszewska']{J. Borowska-Naguszewska}
\affiliation{Institut f\"ur Physik, Humboldt-Universit\"at zu Berlin, Newtonstr. 15, D 12489 Berlin, Germany}
\email{jowita.borowska@desy.de}

\author[orcid=0000-0001-8991-7744, gname='V. Fallah', sname=Ramazani]{V. Fallah Ramazani}
\affiliation{Ruhr University Bochum, Faculty of Physics and Astronomy, Astronomical Institute (AIRUB), 44780 Bochum, Germany}
\email{vandad.fallahramazani@cta-consortium.org}

\author[orcid=0000-0003-1892-2356, gname=R., sname=Konno]{R. Konno}
\affiliation{Department of particle physics and astrophysics, Weizmann Institute of Science, 76100 Rehovot, Israel}
\email{ruslan.konno@weizmann.ac.il}

\author[orcid=0000-0002-9207-4749, gname=D., sname=K\"usters]{D. K\"usters}
\affiliation{Deutsches Elektronen-Synchrotron DESY, Platanenallee 6, 15738 Zeuthen, Germany}
\email{daniel.kuesters@desy.de}

\author[orcid=0000-0003-3457-9308, gname=R.D., sname=Parsons]{R.D. Parsons}
\affiliation{Institut f\"ur Physik, Humboldt-Universit\"at zu Berlin, Newtonstr. 15, D 12489 Berlin, Germany}
\email{daniel.parsons@physik.hu-berlin.de}

\author[orcid=0000-0002-6977-3146, gname=D., sname=Polishook]{D. Polishook}
\affiliation{Department of particle physics and astrophysics, Weizmann Institute of Science, 76100 Rehovot, Israel}
\email{david.polishook@weizmann.ac.il}

\author[orcid=0000-0003-1387-8915, gname=I., sname=Sadeh]{I. Sadeh}
\affiliation{Deutsches Elektronen-Synchrotron DESY, Platanenallee 6, 15738 Zeuthen, Germany}
\email{iftach.sadeh@desy.de}

\author[orcid=0009-0005-7117-8281, gname=O., sname=Savushkin]{O. Savushkin}
\affiliation{Ruhr University Bochum, Faculty of Physics and Astronomy, Astronomical Institute (AIRUB), 44780 Bochum, Germany}
\email{sharrakka@gmail.com}

\author[orcid=0000-0002-0913-3083, gname=E., sname=Segre]{E. Segre}
\affiliation{Department of particle physics and astrophysics, Weizmann Institute of Science, 76100 Rehovot, Israel}
\email{enrico.segre@weizmann.ac.il}

\author[orcid=0000-0002-4667-6730, gname=N., sname=Strotjohann]{N. Strotjohann}
\affiliation{Department of particle physics and astrophysics, Weizmann Institute of Science, 76100 Rehovot, Israel}
\email{nora.linn.strotjohann@gmail.com}

\author[gname=S., sname=Weimann]{S. Weimann}
\affiliation{Ruhr University Bochum, Faculty of Physics and Astronomy, Astronomical Institute (AIRUB), 44780 Bochum, Germany}
\email{sven.weimann@astro.ruhr-uni-bochum.de}

\correspondingauthor{V. Barbosa Martins, N. Jordana Mitjans, \protect\\ S. Garrappa, A. Franckowiak, E. O. Ofek.}

\begin{abstract}

{Optical polarimetry provides information on the geometry of the emitting region, the magnetic field configuration and the properties of dust in astrophysical sources. Current state-of-the-art instruments typically have a small field of view (FoV), which poses a challenge for conducting wide surveys.}
{We propose the construction of the Large Array Survey Telescope Polarization Node (LAST-P), a wide-field array of optical polarimeters. LAST-P is designed for high-cadence ($\lesssim 1$\,day) polarization monitoring of numerous astrophysical transients, such as the early phases of gamma-ray bursts, supernovae, and novae. Furthermore, LAST-P will facilitate the creation of extensive polarization catalogs for X-ray binaries and white dwarfs, alongside a large FoV study of the interstellar medium.
In survey mode, LAST-P will cover a FoV of 88.8 deg$^2$. With a 15\,x\,1-minute exposure, the instrument will be capable of measuring polarization of sources as faint as Gaia Bp-magnitude $\sim$20.9. The precision on the linear polarization degree (PD) will reach 0.7\%, 1.5\%, and 3.5\% for sources with magnitudes 17, 18, and 19, respectively, for a seeing of 2.7$\arcsec$, air mass of about 1 for observations in dark locations. 
We propose three distinct non-simultaneous survey strategies, among them an active galactic nuclei (AGN) strategy for long-term monitoring of $\sim$200 AGN with $<$1-day cadence.
In this paper, we present the predicted sensitivity of the instrument and outline the various science cases it is designed to explore.}

\end{abstract}



\section{Introduction}
Systematic mapping of both the static and transient sky is responsible for many discoveries in modern astronomy. A large number of instruments provides excellent coverage of the sky in terms of multi-band optical photometry \citep{ZTF,ATLAS,ASAS-SN,Panstarrs,GOTO}. However, the community is currently limited to measure the polarization of the optical emission. Crucial information is encoded in the polarization degree (PD) and the polarization angle (PA), as, for example, the geometry of the emitting region and magnetic fields \citep{HOVATTA2019101541, 2019Galax...7...85Z, galaxies7020046, galaxies9020027}; the nature of the acceleration mechanism \citep{Barnier_2024} and accelerated particles \citep{Zhang_2013}.

Technical advances in telescopes and cameras design allow us to build a modular, cost-efficient telescope system \citep{Ofek_2020}. One such example is the Large Array Survey Telescope \citep[LAST, ][]{LASTScience, LAST_pipeline} installed at the Weizmann Astrophysical Observatory (WAO). The first LAST node will include 72 (currently 40 are operational) telescopes  with a large field-of-view (FoV) of $2.2^\circ \times 3.4^\circ \approx 7.4$\,deg$^2$, totalizing a FoV of 532.8\,deg$^2$ for the entire array. The next logical step is to use the multiple small telescopes approach \citep{Ofek_2020} for polarimetry. Here, we suggest the construction of the Large Array Survey Telescope Polarization Node (LAST-P), a new survey telescope for polarimetry. With four polarization filters installed across four co-mounted telescopes observing the same FoV, we will be able to determine the linear PD and polarization PA of all sources in the field. Similar to traditional single-dish telescopes relying on dual \citep{1967PASP...79..136A} or quadruple beam splitting \citep[e.g.][]{Robopol}, LAST-P will simultaneously derive the Stokes parameters without the need to rotate filters, thereby suppressing systematic errors caused by atmospheric or target variability. Uniquely, however, our design offers a much larger FoV, allowing for surveys and monitoring programs to be conducted with high cadence.


We propose an array of 48 telescopes on 12 mounts, with four telescopes per mount in a configuration similar to the LAST array. With each mount observing a $7.4\,\text{deg}^2$ FoV, the complete LAST-P system will achieve a combined instantaneous FoV of $88.8\,\text{deg}^2$. This FoV is orders of magnitude larger than that of existing polarimetric instruments, such as RoboPol \citep[$\approx 0.052\,\text{deg}^2$; ][]{RoboPol2014}, the Nordic Optical Telescope's ALFOSC instrument \citep[$\approx$\,0.011\,$\text{deg}^2$; ][]{ALFOSC2009}, and the Wide Area Linear Optical Polarimeter \citep[$\approx$\,0.25\,$\text{deg}^2$;][]{2024JATIS..10d4005K}. The large FoV of LAST-P allows for the simultaneous monitoring of large celestial patches. This not only facilitates the high-cadence (e.g., daily) observation of numerous sources, but also significantly enhances our ability to detect transient events in both flux and polarization, including those on the fastest timescales \citep{fast_rotations}.

This paper explores the science cases enabled by this new instrument, serving as the primary motivation for its development and construction. We also report on the current status of the project and discuss the path forward. We begin by describing the current design of the instrument, and the estimated capabilities of the future LAST-P in Sect.~\ref{sect:instrument}. Section~\ref{sect:science} delves into the extragalactic and Galactic science cases that LAST-P will enable, focusing on the transient polarimetric sky. Finally, Sect.~\ref{sect:conclusions} outlines the prospects for future developments in the coming years.

\section{Instrument and methods}
\label{sect:instrument}

\subsection{Experimental setup}
LAST is a high-cadence imaging survey, currently comprised of 40 out of a planned 72 telescopes. Each telescope combines a Celestron $27.9\,\text{cm}$ F/2.2 Rowe-Ackermann Schmidt Astrograph (RASA) with a 61-megapixel CMOS sensor, providing a wide $7.4\,\text{deg}^2$ FoV. The system has a plate scale of $1.25\arcsec$ per pixel, and its on-sky image quality is typically $\approx$1.6$\arcsec$ at the field center, degrading to $\sim$3$\arcsec$ -- 4$\arcsec$ at the corners \citep{Ofek_2023}.

With science cases focused on the transient sky \citep{LASTScience}, LAST operates as a discovery machine, utilizing a real-time pipeline for autonomous detection of new sources \citep{LAST_pipeline}, handling a very high (3.3\,Gbit\/s) data production rate. The system's effectiveness has been demonstrated through early results, including the study of the fast blue optical transient AT2022tsd \citep{2025arXiv250818359O}, analysis of the ejecta from the DART-Dimorphos collision \citep{2024MNRAS.52710507O}, and numerous community alerts on new transient detections \citep[e.g.,][]{atel16407, atel16700, atel16729, atel16865}.

The LAST-P design envisions that the four telescopes on each mount will be equipped with a 1.25\,inch Baader Bessel R-filter (from the UBVRI set) and a linear polarization filter. To evaluate the polarizer's efficiency, we use the $25.0\,\text{mm}$ Thorlabs LPVISC100 as a proxy. This component features a high extinction ratio---a measure of its ability to block light polarized perpendicular ($90^\circ$) to the filter's axis. This ratio exceeds $10^4$ across our entire range of interest ($550\,\text{nm}$--$800\,\text{nm}$) and surpasses $10^6$ in the $550\,\text{nm}$ to $650\,\text{nm}$ wavelength range \citep{thorlabs_lpvisc100}. These polarizers, which consist of prolate ellipsoid nanoparticles embedded in sodium-silicate glass, were fixed at relative orientations of $0^\circ$, $45^\circ$, $90^\circ$, and $135^\circ$. Mechanical tolerances in the adapter introduce an alignment uncertainty of several degrees, causing systematic errors in both the PA and the PD. These instrumental effects are corrected by calibrating against a high-polarization standard. In contrast, the camera-to-telescope alignment achieves significantly higher precision, as it is actively determined for each image via astrometric plate-solving.

To validate the LAST-P concept, we implemented a prototype utilizing a single telescope mount from the existing LAST array. Each of the four telescopes was equipped with an R-filter and a linear polarizing filter, fixed at relative orientation angles of $0^\circ$, $45^\circ$, $90^\circ$, and $135^\circ$. This configuration facilitates the intensity measurements necessary to determine the Stokes parameters \citep[$I, Q, U$;][]{serkowski1962}, from which the linear PD and PA of the sources are derived. Detailed results from this test setup will be presented in a forthcoming instrument performance publication, following design refinements aimed at mitigating and accurately characterizing systematic uncertainties. In this work, we focus on estimates derived from the performance of LAST \citep{LAST_pipeline} and the transmittance calculations presented below.

\subsection{Transmittance}
\label{sec:transmittance}
We analyzed the throughput characteristics of LAST-P to evaluate the impact of the color and polarization filters. While a method to derive the effective transmission of the LAST system has been already discussed thoroughly in \citet{garrappa}, this section is based on our laboratory measurements. We used a flux calibration setup at DESY (Zeuthen, Germany), as described in \citet{daniel} to characterize the filters and sensors identical to those used in our instrument setup.

\begin{figure}
    \centering
    \includegraphics[width=\linewidth]{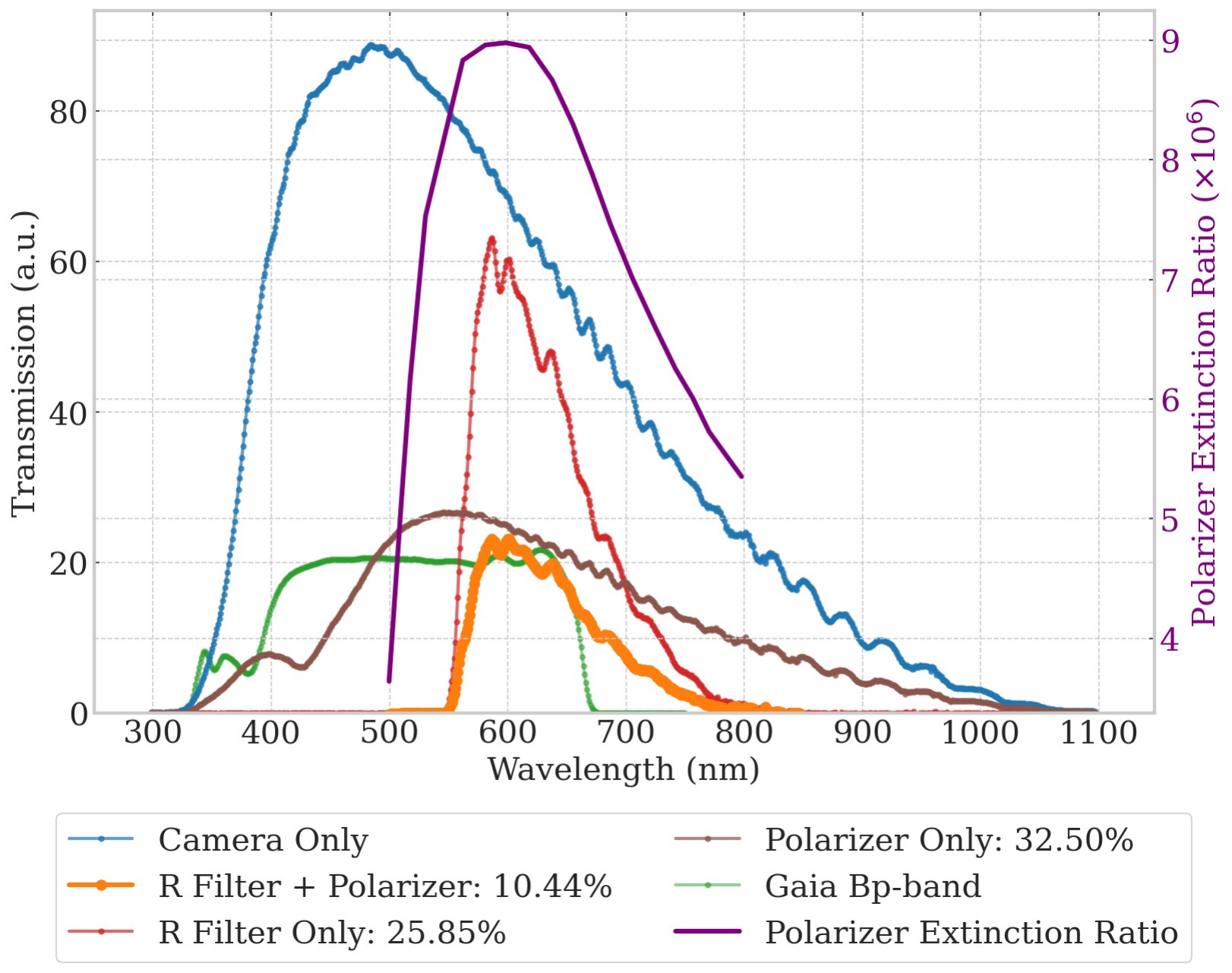}
    \caption{System throughput as function of the wavelength for various instrumental configurations. Configurations plotted include: 'Camera only' (bare sensor QE), 'R Filter + Polarizer', 'R Filter Only', and 'Polarizer Only'. The legend for filter configurations displays their transmittance, representing the system's lambda-weighted throughput relative to the `Camera only' baseline. The `Gaia $Bp$-band' curve shows its "transmission" normalized to the peak of the `R filter + Polarizer Only' system. On the right axis we show the polarizer extinction ratio (purple curve), obtained from \citet{thorlabs_lpvisc100}, showing an excellent performance in the region of interest.}
\label{fig:transmittance} 
\end{figure}

Figure~\ref{fig:transmittance} displays the system's transmittance as a function of wavelength for four distinct configurations: `Camera only,' `R Filter + Polarizer,' `R Filter Only,' and `Polarizer Only.'

To quantify the effective throughput for each configuration, we calculated a lambda-weighted integral of the processed QE curves ($T(\lambda)$), following the methodology in \citet{garrappa} (see Eq.~17): $\int T(\lambda)\,\lambda\,d\lambda$. The transmission factor (transmittance) for each filter setup was then determined by dividing its lambda-weighted integral by that of the `Camera only' configuration. These transmittance, quoted in the legend of Figure~\ref{fig:transmittance}, quantify the throughput relative to the bare camera system.

Our estimates indicate that incorporating the polarization filter reduces the system's lambda-weighted throughput to $\approx$33\% of the `Camera only' value. When both the R-filter and the polarization filter are used, the throughput is further reduced to $\approx$10\%. While atmospheric Rayleigh scattering reduces throughput, its effect is most significant at blue wavelengths. Given that our instrumental design employs an R-filter, which largely blocks this spectral region, the impact on our measurements is small and has therefore not been included in this analysis. For all estimates in this work, we use the Gaia Bp magnitude to maintain consistency with our previous work \cite{LAST_pipeline}.

\subsection{Estimated photometric capability of LAST-P}
\label{sec:instr_photometry}

The LAST system achieves a $5\sigma$ limiting magnitude of $m_{\rm lim,0} \approx 19.6$\,mag (Gaia $B_p$) in a single $t_0 = 20$\,s exposure \citep{LAST_pipeline}. To derive the sensitivity of the new setup with the R-filter and polarization filter, we incorporate the measured noise sources into a unified signal-to-noise ratio (SNR) model. This methodology, fully detailed in Appendix~\ref{app:polarization}, anchors the instrument's performance to the measured $m_{\rm lim,0}$ and it explicitly accounts for both detector properties and the total $\approx10\%$ throughput reduction due to the added R-filter and polarization optics. For an observation of 15 exposures of 60 seconds each ($t_{\rm total} = 900\,{\rm s}$), the projected limiting magnitude is estimated to be 20.4. A darker site would result in lower background noise, yielding an anticipated 50-60\% enhancement in the SNR. This 50-60\% enhancement corresponds to an increase of approximately 0.5 in the dark sky brightness, and hence, on the limiting magnitude. The following estimates (also summarized in Table~\ref{tab:pol_precision}) use these improved dark sky conditions in the extrapolations, since the location of LAST-P is not yet defined but will be preferentially in a darker location. This yields a projected $5\sigma$ detection limit of $m_{\rm lim} \approx 20.9$\,mag for the co-added sequence for the same observing conditions at the LAST site. The limiting magnitude worsens to $m_{\rm lim} \approx 19.4$\,mag ($m_{\rm lim} \approx 19.0$\,mag) for a shorter exposure of $T_{\rm total}=60$\,s ($T_{\rm total}=30$\,s). We caution, however, that our estimates include systematic uncertainties for photometry but not yet for polarimetry. A detailed study of instrumental polarization requires the finalized instrument design and is planned for a future stage.

\subsection{Estimated polarimetric capability of LAST-P}
\label{sec:instr_polarimetry}

To assess the polarimetric potential of LAST-P, we map the photometric SNR to the expected uncertainties in PD, PA, and the Minimum Detectable Polarization ($MDP$).

While the exact reduction of polarimetric data involves error propagation (detailed in Appendix~\ref{app:polarization}), we utilize the simplified approximations derived in Appendix~\ref{app:sensitivity_approx} to estimate the instrument's theoretical sensitivity. The absolute uncertainty in the $PD$ scales inversely with the $SNR$:
\begin{equation}
    \label{eq:pd_main_text}
    \sigma_{PD} \approx \frac{1}{SNR}
\end{equation}
Similarly, the $PA$ uncertainty depends on the source's intrinsic polarization:
\begin{equation}
    \label{eq:pa_main_text}
    \sigma_{PA} [^\circ] \approx \frac{28.65^\circ}{PD \cdot SNR}
\end{equation}

We also calculate the $MDP$ in percentage at the 99\% confidence level (see Appendix~\ref{app:mdp_def}):

\begin{equation}
    \mathrm{MDP} \approx 368.4 \times \frac{1}{\mathrm{SNR}}.
    \label{eq:mdp_main_text}
\end{equation}

Table~\ref{tab:pol_precision} summarizes these capabilities for a $15 \times 60$\,s observing sequence. These estimates rely on the derived $5\sigma$ limiting magnitude of $m_{\rm lim} \approx 20.9$, assuming a representative delivered image quality (FWHM) of 2.7\arcsec~\citep{LAST_pipeline}, and using the SNR scaling developed in Appendix~\ref{app:polarization}. For estimates with different exposure times and source magnitudes we refer the reader to Appendix~\ref{app:complete_table}.

\begin{table*}[t]
    \centering
    \caption{Estimated Polarimetric Precision for the Survey Mode.}    \label{tab:pol_precision}
    \begin{tabular}{cccccccc}
    \hline
    \textbf{Mag} & \textbf{SNR} & \textbf{$\sigma_{PD}$} & \textbf{MDP (99\%)} & \multicolumn{3}{c}{\textbf{$\sigma_{PA}$ ($^\circ$)}} \\
    \cline{5-7}
     & & (absolute \%) & (\%) & \textbf{$PD=5\%$} & \textbf{$PD=10\%$} & \textbf{$PD=15\%$} \\
    \hline
    17.0 & 143 & 0.7 & 2.6 & 4.0 & 2.0 & 1.2 \\
    18.0 & 66  & 1.5 & 5.6 & 8.7 & 4.3 & 2.9  \\
    19.0 & 28 & 3.5 & 13.1 & 20.3 & 10.2 & 6.8 \\
    \hline
    \end{tabular}
    \par\medskip
    \small
    \emph{Note:} Estimates derived from a limiting magnitude of $m_{\rm lim} \approx 20.9 \text{ mag}$ for a $15 \times 60 \text{ s}$ co-added observation. For the Deep Mode of observation, the estimates improve by a factor of 2. $\sigma_{PA}$ values are based on the small-angle approximation. MDP is calculated at the 99\% confidence level from Eq.~\ref{eq:mdp_approx}. Additional estimates can be found in Table~\ref{tab:complete_table_updated}.
\end{table*}

We emphasize that these estimates reflect the statistical photon-noise limit; in practice, the achievable precision may be further constrained by systematic uncertainties, such as instrumental polarization residuals and sky subtraction errors. 

\subsection{Survey Strategy}
\label{sec:survey}
The main advantage of our instrument is its wide FoV, which enables polarization monitoring of a large number of sources with high cadence. Following the standard LAST photometric observing strategy \citep{LASTScience}, we plan to operate three distinct non-simultaneous strategies: a \textit{high-cadence} survey, a \textit{low-cadence} survey, and an \textit{AGN} strategy.

In the high-cadence survey, the same sky regions will be visited three times per night. Depending on source density and on the scientific interest of each region, we will adopt either the survey mode (with one unit centered on each field) or the deep mode (with four units co-pointed for increased sensitivity).

We assume an average of 5 hours per night dedicated to survey observations, excluding calibration and follow-up time for Targets of Opportunity. With three visits per region, 15 observations per visit, and 1 minute per observation, this corresponds to a total of $3 \times 15 \times 1\,\text{min} = 45\,\text{min per region}$. This allows coverage of up to 6 regions per night (accounting for overheads due to re-pointing).
The instrument has a FoV of $88.8\,\text{deg}^2$. We therefore cover $6 \times 88.8 = 532.8\,\text{deg}^2$ per night in survey mode, and  $532.8 / 4 = 133.2\,\text{deg}^2$ per night in deep mode (four telescopes observing the same region).

Given that we consider the same geographical latitude as for the LAST site, sources can be observed on average down to a declination of $-30^\circ$, corresponding to a total accessible sky area of $A \approx 31,000\,\text{deg}^2$. This represents  75\% of the full available sky, not accounting for limitations due to constraints on the maximum allowed air mass. Therefore, approximately 58 nights are required to cover the entire visible sky at least once in the high-cadence survey mode. Depending on our science goals, it may be beneficial to repeatedly observe the same sky-region over an extended period, e.g., a full week. This strategy is particularly valuable during the early phases of the survey; it enables validation of our measurements, as well as cross-correlation of variability on timescales longer than a day with data from existing instruments.

In the low-cadence survey, our goal is to cover the largest possible area of the sky as quickly as possible, while still using the standard sequence of 15\,x\,1-minute exposures. To survey the full available sky area of $\approx 31,000\,\text{deg}^2$ using a FoV of $88.8\,\text{deg}^2$, we require 349 distinct pointings. This corresponds to a total of 87.2 hours of observation. Assuming 5 hours of observing time per night for the survey, the entire region can be covered in approximately 18 nights. This estimates accounts for time lost to re-pointing between fields to complete a full-sky pass, considering the available area of about $31,000\,\text{deg}^2$. 

A third survey strategy, focused on bright AGN, will also be implemented. As part of the AGN strategy, each field is observed in survey mode three times over the same night, using 1-minute exposures per-visit. This approach enables coverage of 120 distinct regions per night, each spanning $88.8\,\text{deg}^2$. It results in a total nightly coverage of approximately $10,\!656\,\text{deg}^2$, or about 34\% of the accessible sky. Under this strategy, an estimated number of 200 AGNs from the GaiaunWISE catalog \citep{gaia_unwise_agn} with Gaia Bp-band magnitude brighter than 16 could be monitored with cadence of less than a day (See Sect.~\ref{sec:agn} for details on AGN science). The precision on the PD (PA) of the daily monitored sources with 1-minute exposures would be as low $\approx$1.3\% (3.8$^\circ$) for 16 magnitude sources with a PD of 10\% (see Table~\ref{tab:complete_table_updated} for more estimates). 

We emphasize that these estimates do not take into account seasonal variations in sky visibility or fluctuations in the total number of observing hours available per night. Rather, they serve as the baseline for developing optimized survey strategies. Table~\ref{tab:survey_summary} summarizes the survey strategies and modes discussed in this section.

\begin{table*}[htbp]
\centering
\caption{Summary of sky coverage and time estimates for each observing mode and strategy.}
\label{tab:survey_summary}
\begin{tabular}{cccc}
\hline\hline
\textbf{\thead{Strategy}} & \textbf{\thead{Mode}} & \textbf{\thead{Sky Area\\per Night}} & \textbf{\thead{Full Available\\Sky in}} \\
\hline
High-Cadence & Survey & $532.8\,\text{deg}^2$ & 58 days \\
High-Cadence & Deep   & $133.2\,\text{deg}^2$ & 233 days \\
Low-Cadence  & Survey & $1,\!721.5\,\text{deg}^2$ & 18 days \\
AGN  & Survey & $10,\!656\,\text{deg}^2$ & -- \\
\hline
\end{tabular}
\end{table*}

In addition to conducting a continuous survey and responding to community alerts, LAST-P will issue its own rapid photometric and polarimetric alerts for transient events. A key function of these alerts is to facilitate coordinated optical and gamma-ray observations. LAST-P is also designed to enhance multi-wavelength science by participating in joint campaigns with other observatories. Based on approved proposals, it will perform simultaneous observations with high-energy facilities such as the Cherenkov Telescope Array Observatory \citep[CTAO, ][]{cta2018science} and the Imaging X-ray Polarimetry Explorer \citep[IXPE, ][]{WEISSKOPF20161179}. These distinct operational modes support the diverse science cases detailed below.

\section{Science with LAST-P}
\label{sect:science}
This section outlines the primary extragalactic and Galactic science cases that LAST-P will address.

\subsection{Extragalactic Science}
\label{sect:extra}
\subsubsection{AGN}
\label{sec:agn}

The linear polarization of AGN, particularly blazar jets, provides a powerful diagnostic of jet magnetic-field geometry, particle acceleration, and flow dynamics. Optical synchrotron emission can originate from the compact core, in moving plasmoids or shocks, or within turbulent jet plasma \citep{Marscher}. Contemporaneous polarization observations of blazars in the optical, as well as in the X-ray band, now available due to IXPE, reveal a wavelength-dependent, stratified shock structure where the PD systematically decreases toward longer wavelengths \citep{nature_ixpe, ixpe}. Multi-zone models further predict higher-amplitude variations for the X-ray signal compared to longer wavelengths, together with random rotations of the PA, consistent with a moving shock scenario interacting with turbulent magnetic fields. In turbulent environments, randomly oriented zones compete and their polarization vectors partially cancel each other. Smoother and more systematic PA rotations may indicate coherent jet structures, such as a knot propagating along a helical magnetic field. The dominant orientation of the PA rotations may be linked to the rotational direction of the ergosphere of the black hole or of the accretion disk \citep{Shishkina}.

Changing-look active galactic nuclei (CL-AGN) are objects whose broad Balmer emission lines appear or disappear on timescales of months to years. This behavior is explained by two competing models, variable obscuration and changes in the accretion state. The obscuration model proposes that a dusty cloud moves into our line of sight, blocking the nucleus. In this case, only highly polarized, or infrared, scattered light would be visible; the "hidden" broad lines should still be detectable in this polarized spectrum \citep[e.g., the $\sim$16\% polarized scattered light from NGC 1068; ][]{Antonucci1985}. Conversely, the changing-state model suggests the engine itself dims due to a drop in the supermassive black hole accretion rate. This intrinsic change would stop the production of broad lines and result in a very low degree of polarization. Observational evidence favors the latter model, since CL-AGN have consistently been found to have low polarization \citep{changing_look_agn_pol, changing_agn_nature}. This suggests that changes in the accretion rate are the primary driver for most such events. However, the rapid timescales of these changes are more consistent with disk instabilities than with the orbital periods of torus clouds. Such instabilities may even temporarily launch short-lived radio jets in some CL-AGN, further reinforcing the transient accretion nature of their behavior \citep{birmingham2025birthyoungradiojets}. The LAST-P survey will significantly expand the sample of changing-look AGN with polarization measurements. This will enable a robust population study to better distinguish between competing models and clarify the primary cause of this spectral variability.

Polarization measurements of AGN also help constrain acceleration mechanisms. In highly magnetized AGN jets, magnetic reconnection is likely to dominate over shock acceleration. In reconnection scenarios, the PD in the X-ray regime is expected to be lower than in the optical, with a distinct temporal evolution compared to shock-driven processes \citep{2022A&A...662A..83D}. In addition, an extremely high PD ($>$40\%) in the optical regime during a flare, contrasted with a much lower PD in the X-ray band ($<$10\%), may help break the degeneracy between hadronic and leptonic models, favoring the latter. This has e.g., been argued for BL Lac during an observation campaign in November 2023 \citep{Agudo_2025}.

A further key aspect to investigate is the reported temporal association between several GeV flares observed by \textit{Fermi}-LAT and large events exhibiting systematic rotations of the optical PA \citep{10.1093/mnras/stx2786}. This association hints at a common acceleration site for the optical and gamma‑ray–emitting particles in these cases. Such observations are important for gamma-ray analyses of AGN, due to the poor spatial resolution ($\sim$0.1$^\circ$ -- $\sim$1$^\circ$) of gamma-ray instruments. Although this correlation has been observed for several gamma-ray flares \citep{3C279, Sorcia_2014, 10.1093/mnras/stx2786, Shishkina, 2015ApJ...813...51C}, these events do not exhibit distinctive properties that clearly distinguish them from other flares; this leaves the underlying mechanism open for investigation. By observing the same AGN contemporaneously with VHE gamma-ray observations from Imaging Atmospheric Cherenkov Telescopes (IACTs), we aim to test whether VHE gamma-ray flares may also be accompanied by PA rotations.

Large variations in PD and PA on sub-hour timescales have been reported in a high-cadence strategy by \citet{multi_instrument}, which combined data from different instruments. This study, along with others \citep[e.g.,][]{fast_rotations}, underscores the critical need for a high-cadence survey instrument capable of systematically detecting rapid PA rotations. A facility that combines high-cadence, wide-field, and polarimetric monitoring is therefore essential to fully leverage the diagnostic potential of time-resolved AGN polarization.

\begin{figure}[]
\centering
\includegraphics[width=\linewidth]{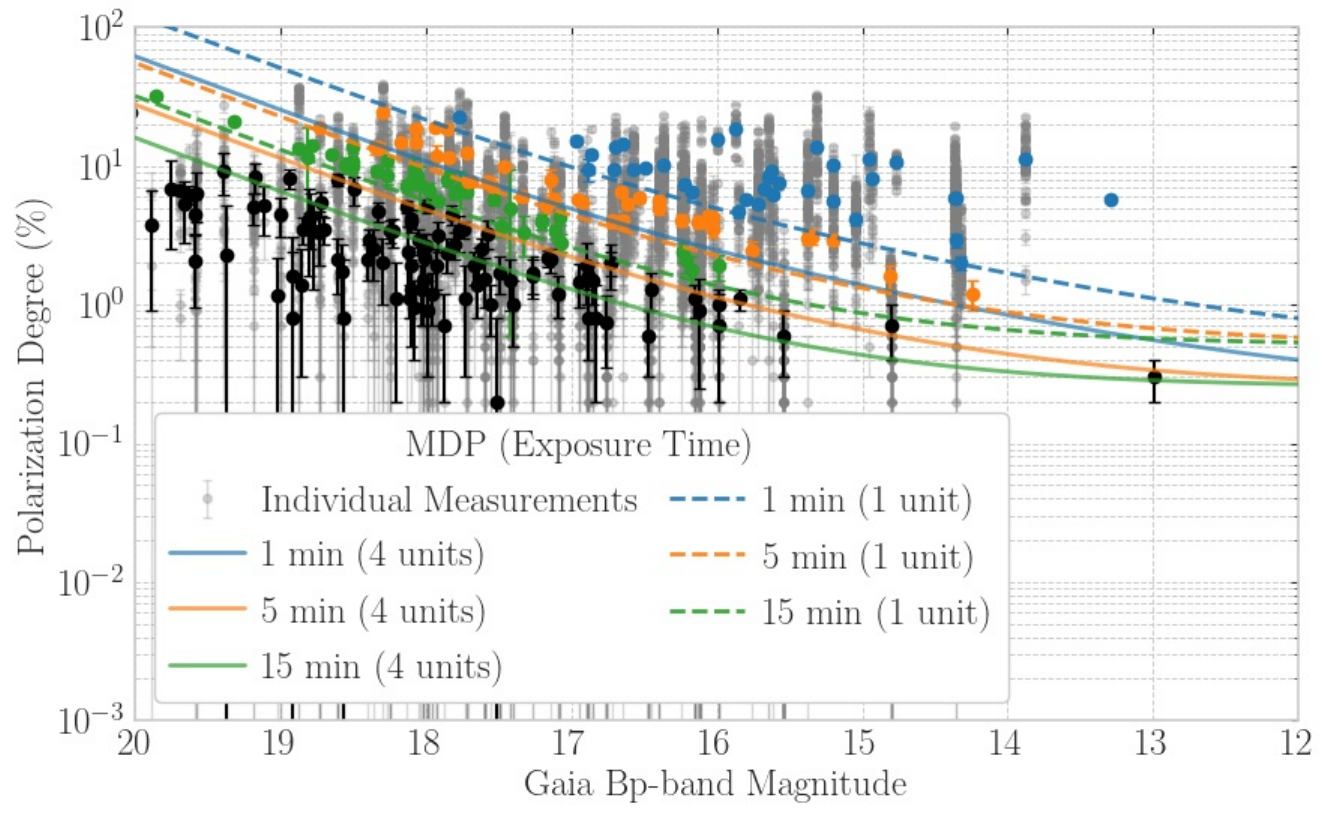}
\caption{Polarization degree of AGN from the RoboPol AGN catalog \citep{robol_cat}. Gray points indicate individual observations, while colored points show the median polarization degree for each source. The color scale represents the minimum exposure time required for our instrument (assuming four telescopes, i.e., one unit) to achieve the sensitivity needed to detect that polarization, while black points represent AGN that fall outside the 15\,x\,1\,min curve for one unit. Colored dashed and solid lines indicate the MDP curves of LAST-P for different exposure times using survey mode (one unit) and deep mode (four units), respectively. Typical seeing of 2.7$\arcsec$ and air mass of approximately 1 are assumed.
}
\label{fig:robopol}
\end{figure}

We use data from the RoboPol AGN catalog \citep{2021MNRAS.501.3715B, robol_cat} cross-correlated to the Gaia DR3 catalog \citep{Gaia_3} to evaluate the monitoring capabilities of our instrument under different observation strategies. Figure~\ref{fig:robopol} presents these data alongside the MDP sensitivity curves of LAST-P for various exposure times. With a typical 15\,$\times$\,1-minute observation, our instrument would be able to monitor up to 116 of the 221 AGN from this catalog. Moreover, exposure times can be adapted based on expectations for where each AGN will fall in the brightness-magnitude phase-space (Fig.~\ref{fig:robopol}). For instance, BL-Lacertae is typically bright ($R$-band AB-magnitude $\sim$12) with PDs around $\sim$15\%, placing them well above the first blue dashed sensitivity curve. 

\begin{figure}[]
\centering
\includegraphics[width=\linewidth]{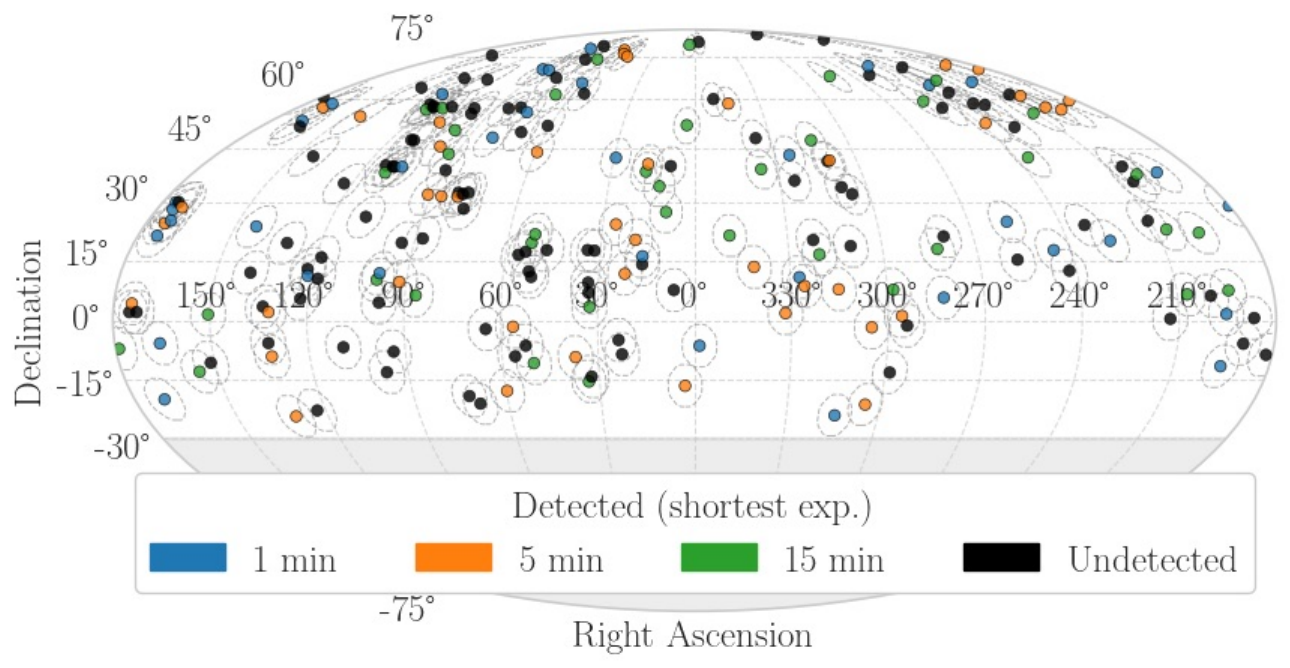}
\caption{Spatial distribution of AGNs observed by RoboPol \citep{robol_cat}. Colored dots represent AGNs, where the color encodes the minimum exposure time needed to measure their PD. Dashed gray circles show the FoV of our instrument ($88.8\,\text{deg}^2$) centered on each AGN. The gray region below declination $-$30$^\circ$ is hardly accessible to LAST-P due to its assumed future geographic location.}
\label{fig:agn_sky_pol}
\end{figure}

The large FoV of our instrument will enable simultaneous monitoring of multiple AGNs within a single observation, particularly in survey mode, which covers 88.8\,deg$^2$ per pointing. Therefore, we will observe and monitor a larger number of AGN than present in the current RoboPol AGN catalog. In fact, based on the \textit{GaiaunWISE} AGN catalog \citep{gaia_unwise_agn}, we identify a total of 2,400 sources with Gaia $Bp$-band magnitudes brighter than 17, an AGN classification probability exceeding 99\%, and declinations above $-30^\circ$. These would be accessible to LAST-P, expected to yield PD uncertainties of 0.7\%, PA uncertainty of 2$^\circ$ (for a PD of 10\%), and MDP of 2.6\% over 15\,x\,1\,minute exposures. For the 60-second exposures (see AGN strategy in Sect.~\ref{sec:survey}), LAST-P will reach PD uncertainty of 1.3\%, PA uncertainties of 3.8$^\circ$ (for a PD of 10\%), and a MDP of 4.9\% for the $\sim$200 \textit{GaiaunWISE} AGN brighter than Bp-magnitude 16. Figure~\ref{fig:agn_sky_pol} displays the spatial distribution of AGNs observed by RoboPol, with the gray circles illustrating the coverage area of our instrument around each source. The color scale represents the minimum exposure time required to measure the PD of each AGN, based on the median values from RoboPol. The sky-region accessible to LAST-P closely matches that of RoboPol, ensuring high complementarity with existing data.

\subsubsection{Gamma-ray Bursts}
Gamma-ray Bursts (GRBs) are considered the most energetic explosions in the Universe. They are mostly discovered by observations of their bright prompt gamma-ray emission using dedicated gamma-ray telescopes, such as the Swift Burst Alert Telescope \citep[BAT; ][]{2005SSRv..120..143B,2016ApJ...829....7L}. Owing to real-time alerts by gamma-ray telescopes, there has been numerous detections of long-lived afterglows after the initial prompt GRB emission at wavelengths across the electromagnetic spectrum \citep[e.g., ][]{1997Natur.387..783C,1997Natur.386..686V,1998ApJ...500L..97G}. Most recently they have also been detected at TeV energies \citep[e.g., ][]{MAGIC_GRB, HESS_GRB, LHAASO_GRB}. 

The characteristic prompt gamma-ray emission of GRBs is produced by highly-relativistic bipolar outflows, as a result of presumably either a merger of compact object binaries \citep[e.g., ][ for spectrally-hard \textit{short GRBs}]{2013Natur.500..547T}, or a collapse of massive stars \citep[e.g., ][ for \textit{long GRBs}]{2003Natur.423..847H}. Short GRBs are interpreted to be associated with gravitational wave emission from compact binary mergers \citep[e.g.,][]{2017ApJ...848L..13A}, and long GRBs are associated with some core-collapse supernovae. However, the majority of detected GRBs lack an observed counterpart (either gravitational wave signal or associated supernova). The afterglow emission of GRBs is interpreted as synchrotron radiation emitted when the jet ejected material is decelerated by the ambient medium by a pair of shocks, i.e., a forward shock that propagates into the circumburst medium \citep[e.g., ][]{2000ApJ...545..807K} and a reverse shock that crosses the ejecta.

\begin{figure}[]
\centering
\includegraphics[width=\columnwidth]{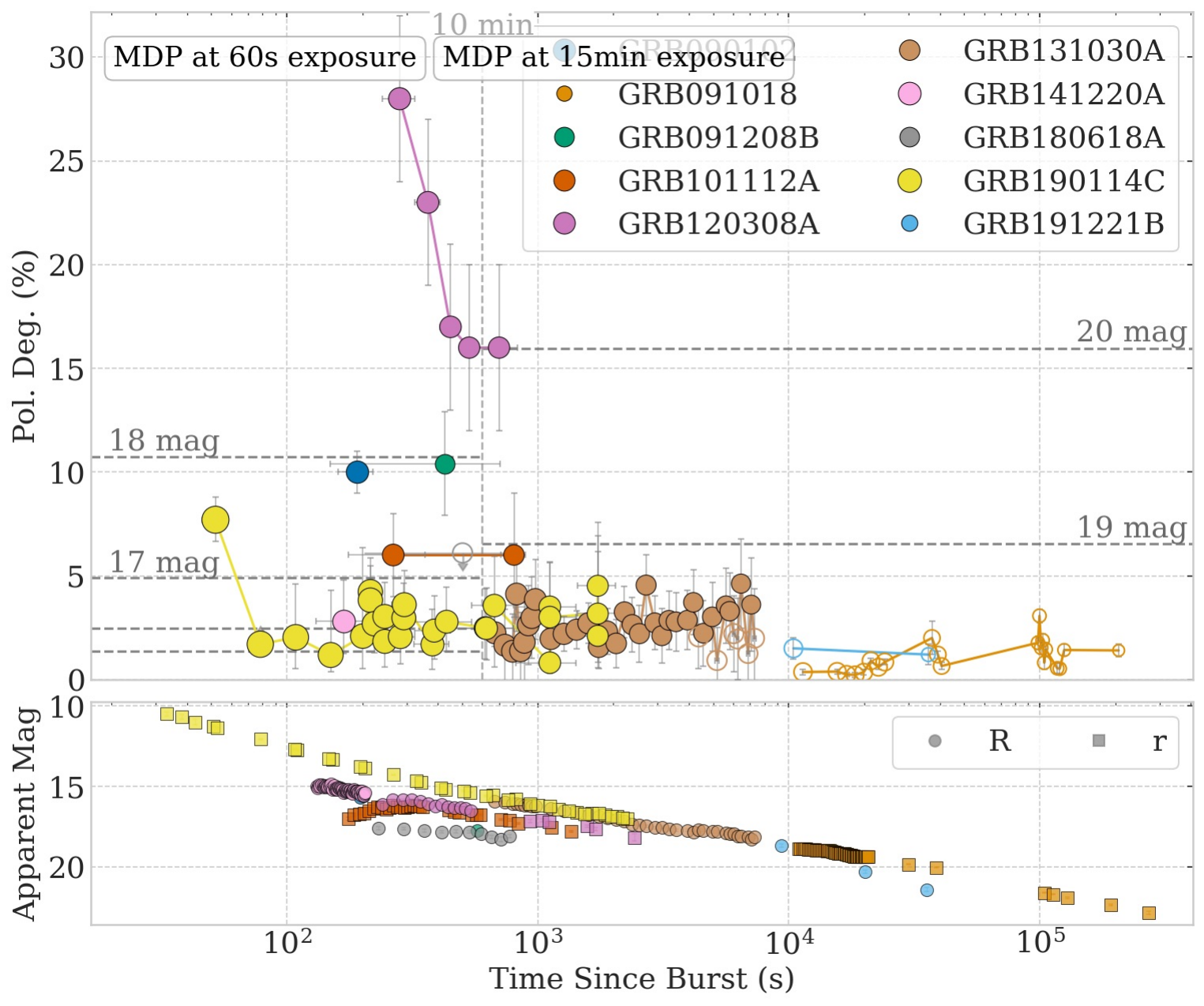}
\caption{Top panel: Evolution of optical afterglow PD for a GRB sample. Data points are sized according to the apparent R-magnitude measurement closest in time, with brighter magnitudes corresponding to larger points. Points falling below the relevant sensitivity threshold for their associated magnitude are shown unfilled. On the left side we consider the sensitivity (MDP) for an exposure of 60 seconds and on the right for an exposure of 15\,x\,1 minute in the deep mode of observation. The dashed lines represent the MDP for sources with magnitudes 18, 17, 16, 15 (in the $<$10 minutes region) and with magnitudes 20 and 19 (in the $>$10 minutes region) from top to bottom. Bottom panel: The corresponding apparent magnitude (\text{R} or \textit{r} according to available data) . Data compiled from: GRB 090102 \citep{2009Natur.462..767S, 10.1111/j.1365-2966.2010.16601.x}, GRB 091018 \citep{2012MNRAS.426....2W}, GRB 091208B \citep{Cano2009GCN10262, 2012ApJ...752L...6U},  GRB 101112A \citep{2017ApJ...843..143S}, GRB 120308A \citep{2013Natur.504..119M, 2017ApJ...843..143S}, GRB 131030A \citep{2014MNRAS.445L.114K}, GRB 141220A \citep{2021MNRAS.505.2662J}, GRB 180618A \citep{Jordana-Mitjans_2022}, GRB 190114C \citep{2020ApJ...892...97J}, GRB 191221B \citep{2021MNRAS.506.4621B, Chen_2024}.}
\label{fig:grb_pol_evolution}
\end{figure}

By measuring the polarization of the short-lived reverse shock emission, it may be possible to distinguish between magnetically dominated \citep{2009MNRAS.394.1182K,2011ApJ...726...90Z} and baryonic jet models \citep{1993ApJ...405..273W,1999ApJ...526..697M}. Significant levels of polarization (PD of 5\%\,--\,30\%) have been detected during the first couple of minutes after the burst (see Fig.~\ref{fig:grb_pol_evolution}), which points to a large-scale ordered magnetic field rather than a pure baryonic jet. High levels of polarization can be explained by the modeling of the reverse shock emission in mildly magnetized jets \citep{2013Natur.504..119M,2014ApJ...785...84J}. However, prompt observations of the TeV-detected GRB 190114C hint that the most energetic jets could be highly magnetized \citep{2020ApJ...892...97J}. A statistical approach with fast-follow-up polarimetry is required to further constrain models in the low PD regime (P\,$<$\,2\%; \citealt{2011ApJ...734...77I}). In addition, forward shock polarimetric measurements allow the study of the angular structure of jets \citep[e.g., ][]{2014Natur.509..201W}, as well as of the dust properties of the host galaxies of GRBs \citep[e.g., ][]{2021MNRAS.505.2662J}. 

A study on the potential of independent discovery of GRBs using LAST-P was performed by \citep{Sadeh:2023ohn}. In the current work, we focus on rapid follow-up observations, given external science alerts.
GRB alerts from the BAT have small uncertainty in the position of the event \citep[i.e. $\approx 1-4\,$arcmin at 90\% confidence; ][]{2005SSRv..120..143B}. This has allowed successful follow-up of GRBs, starting seconds to minutes post-burst, using robotic instruments. In particular robotic facilities with polarimetric capabilities have opened a window into the early evolution of the polarization signal (MASTER; \citealt{2012ExA....33..173K}, Liverpool Telescope; \citealt{2006PASP..118..288G}, Kanata Telescope; \citealt{2008SPIE.7014E..4LK}, Skinakas Observatory; \citealt{2014MNRAS.442.1706K}). Given that \textit{Swift}/BAT detects an average of 88 GRBs per year, with around 51\% exhibiting detectable optical afterglows\footnote{\url{https://swift.gsfc.nasa.gov/archive/grb_table/stats/}}, we expect to follow up and detect roughly 9 GRBs annually with LAST-P. This includes potential detections during the prompt emission phase, depending on the timing of the alerts and the response time of our planned automatic follow-up system.


When the relative angle between the line of sight of the observer and the jet of the GRB is larger than the relativistic beaming angle of the GRB, the characteristic prompt gamma-ray emission is very weak, most likely undetectable. 
The propagation of the relativistic jet through the stellar envelope is expected to inject energy into the stellar shocked material, producing cocoon emission \citep{2017ApJ...834...28N}. For such scenarios, low and rather constant polarization is expected \citep{2018MNRAS.478.4128G,2018ApJ...867...18N}. 

LAST-P will measure GRBs during $t\approx 0.5-10 \,$min post-burst, allowing to explore a regime with sparse polarimetric data (see Fig.~\ref{fig:grb_pol_evolution}). The deep mode will be employed to allow us to detect faint GRBs and/or achieve good PD statistics for a reduced exposure time in the prompt phase. The exposure time will be automatically adjusted from $5\,$s up to 60\,s, as  $53\%$ of the GRBs detected by BAT have a duration of $\gtrsim 35\,$s \citep{2016ApJ...829....7L}. The individual observations would later be stacked, increasing the sensitivity for fainter afterglows. Polarization detected during forward shock emission has been systematically lower than that of the early-time reverse shock emission. Therefore, the observing benchmark would be $15$-min exposures for $\gtrsim 10 \,$min post-burst. Correspondingly, we expect to achieve polarimetric precision of $\approx 1.8\%$ for sources brighter than $19\,$mag, using the deep mode of observation. Furthermore, the large FoV and high cadence of LAST-P will allow a confident mapping and subtraction of the polarization induced by Galactic dust in the line of sight, using large statistics of field stars. 

Figure~\ref{fig:grb_pol_evolution} presents a compilation of PD and apparent $R$-band magnitude measurements for several GRB afterglows as a function of time post-trigger. These observations are compared to the estimated sensitivity of LAST-P for a total exposure of 1 minute over a period of up to 10 minutes in deep mode. We note that the sensitivity of LAST-P is estimated in the $G$-band. 
%
While all early-time measurements fall within the expected reach of LAST-P, observations become challenging as the afterglow fade, particularly for events reaching magnitudes as low as 20. With a 30-second exposure during the early stages of a GRB (within the first 10 minutes), LAST-P would be able to resolve variations in the PD as low as 3.8\% (1.9\%) at magnitude 17 (16) in survey mode.

Complementing alerts from BAT and GBM (subject to their operational status), we anticipate following up on triggers from the recently launched Space-based multiband astronomical Variable Objects Monitor mission (SVOM; \citealt{2016arXiv161006892W}), the \textit{Einstein Probe Mission} \citep{yuan2022einstein}, and the Vera C. Rubin Observatory’s LSST \citep{LSST}. Overall, the fast cadence  of LAST-P will provide new insights into the early optical emission of GRBs. We expect the instrument to contribute invaluable data for multiwavelength and multi-messenger analyses of GRBs.

\subsubsection{Supernovae}
Instruments such as LAST-P, with its high cadence monitoring survey, have the potential to perform early-stage SN polarimetry. Measuring intrinsic SN polarization is complicated by foreground contamination from the interstellar medium (ISM) in the Milky Way and as part of the host galaxy. While some methods exist to partially correct for such effects \citep{ISM_pol_method}, continuous monitoring can be crucial. For instance, it can be used to separate the time-variable SN polarization from the relatively constant galactic ISM component, especially when high-cadence observations are available.

Polarimetry of SN types probes the geometry of the event \citep[e.g., ][]{2022MNRAS.511.2994L}. Type Ia SNe typically show low broadband continuum polarization. In addition, they can exhibit strong line polarization before peak luminosity, which may be linked to the composition of the ejecta and to the viewing angle. Core-collapse SNe (Types II, Ib, Ic) show evidence for asphericity, often suggesting asymmetries aligned along a specific axis (e.g., the existence of a jet). In Type II SNe, the PD is often modest initially, but can increase significantly (to $\approx$1\% or more) at later times (tens to hundreds of days post-explosion); e.g., as the photosphere recedes through the hydrogen envelope, the aspherical inner core is revealed \citep[see e.g. for a review, ][]{Wang2008}. 

\begin{figure}
    \centering
    \includegraphics[width=\linewidth]{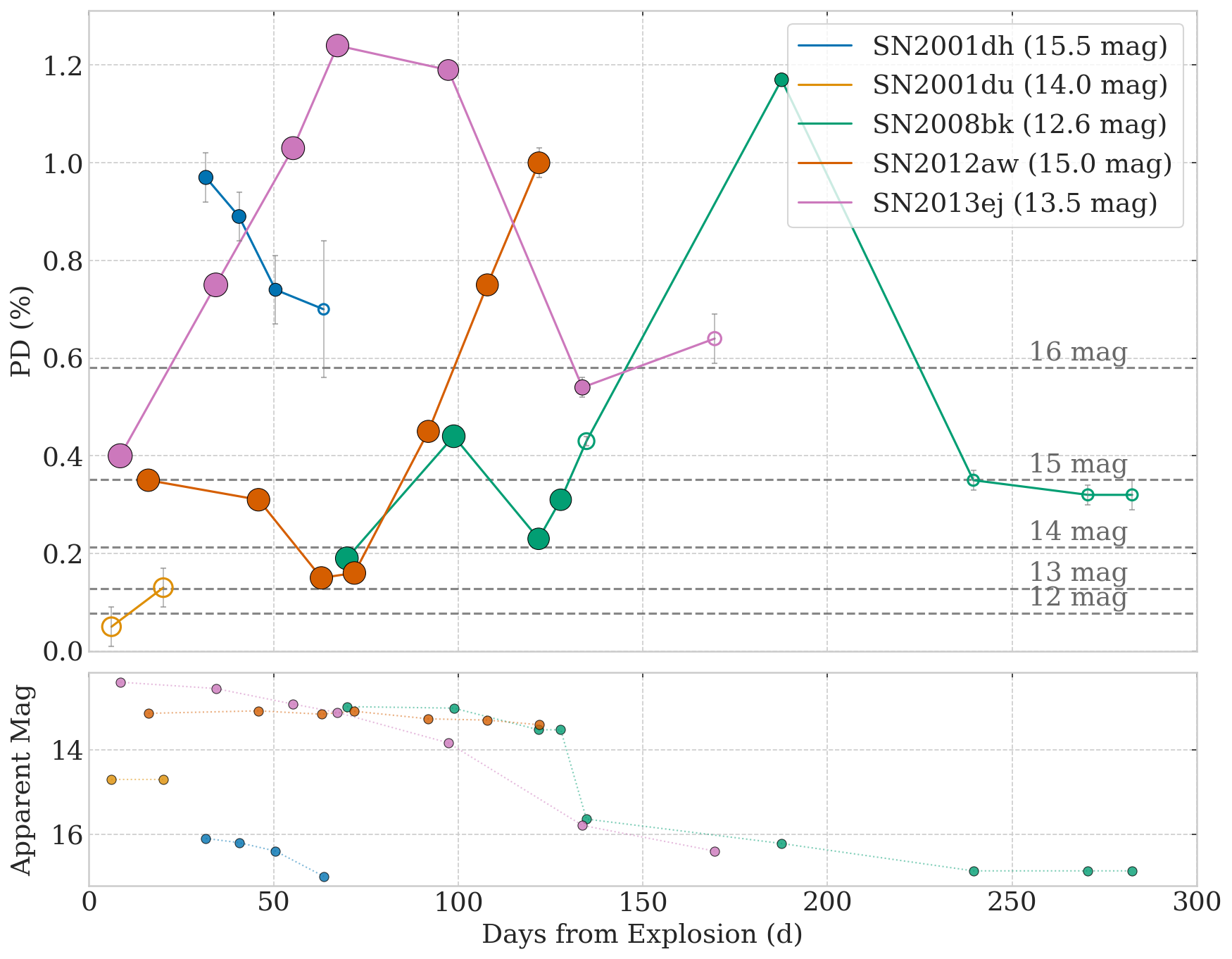}
    \caption{Top panel: PD as function of time for the supernova sample \citep{Nagao2024}. Data points are sized according to the apparent magnitude measurement closest in time extracted from \citet[][]{AstroCats_Software}, with brighter magnitudes corresponding to larger points. Points falling below the relevant sensitivity threshold for their associated magnitude are shown partially transparent. Dashed lines represent the MDP for the a 15\,x\,1 minute exposure in deep mode of observation for sources with magnitudes 16, 15, 14, 13, 12 from top to bottom. Bottom panel: Apparent $R$-band magnitude evolution is shown for all objects, except for SN2008bk, for which the $V$-band magnitude is plotted.}
\label{fig:sn_pol_evolution} 
\label{fig:wd_plots} 
\end{figure}

Figure~\ref{fig:sn_pol_evolution} shows the magnitude and PD for several Type II SNe \citep{Nagao2024}, given the relative sensitivity of LAST-P in the G band. The  increase in the PD generally correlates with decreasing brightness. This was first observed in SN 1987A, where the PA remained constant, indicating a stable asymmetry axis \citep{Jeffery1991}. Core-collapse SNe also often exhibit loops in the Stokes Q-U plane, due to wavelength-dependent polarization variations linked to ejecta structure. LAST-P can potentially observe PD changes in most of the SN shown in the figure. As the PDs of supernovae are expected to be relatively small, multiple stacked exposures would be necessery.

Type Ib/c SNe (lacking outer hydrogen/helium) tend to show higher polarization than Type II SNe (e.g., SN 1997X reached PD $\approx$ 7\%), suggesting significant explosion asymmetries \citep{Wang2001}. While sharing common features, peculiar SNe like the double-peaked SN 2005bf \citep{Tominaga_2005} benefit from polarimetry for classification and physical understanding. The link between long-duration GRBs and energetic, aspherical core-collapse SNe has been established. While most core-collapse SNe have no detected GRB counterpart, some bright SNe may be linked to quite dim GRBs \citep[e.g., SN 1998bw, ][]{1999ApJ...516..788W}. Asymmetries can arise from jets, accretion outflows, asymmetric neutrino emission, magnetorotational effects, or clumpy ejecta. Therefore, polarimetric SN observations are powerful diagnostics. They can be used to probe explosion mechanisms; the SN-GRB connection; and interstellar dust properties in host galaxies via foreground polarization characterization. LAST-P will complement the All-Sky Automated Survey for Supernovae \citep[ASAS-SN, ][]{10.1093/mnras/stac3801} with its approximately 250 SNe brighter than G peak magnitude 16 detected or recovered from 2018 to 2020 \citep{10.1093/mnras/stad355}. For Gaia Bp magnitude of 16, LAST-P will be able to detect polarization (MDP) down to 0.7\% using deep mode observations in a 15\,x\,1 minute exposure.

\subsubsection{Tidal Disruption Events}

Tidal disruption events (TDEs) occur when a stellar object is torn apart by the tidal forces of a supermassive black hole at the center of a galaxy. So far, multi-wavelength observations and polarimetry of TDEs have proven useful to provide some hints regarding jet physics in supermassive black holes \citep[e.g., ][]{floris2025polarimetricdiversitytidaldisruption, jordanamitjans2025opticalpolarizationstellarfedactive}. They have allowed the discovery of a few relativistic jets in TDEs among the general population of thermal TDEs \citep[e.g., ][]{2011Sci...333..203B,2019MNRAS.482.5023H,2020MNRAS.491.1771W,2020ApJ...892L...1L}. In particular, the detection of variable polarization in Swift J2058+0516 (PD of $< 5.3\%$ to $8.1 \pm 2.5 \%$ ; \citealt{2020MNRAS.491.1771W}) and AT2019dsg ($P= 9.2 \pm 2.7\%$ to $P=2\%$; \citealt{2020ApJ...892L...1L}) suggested the presence of a relativistic outflow. \citet{tde_3_measurements} shows the PD of three TDEs decreasing since the time of detection, as well as a continuum PD that is independent of the wavelength, suggesting electron scattering as the dominant source of polarization in TDEs. The shock generated by the collision of the stellar debris stream with itself has arisen as an additional explanation for the extremely high host-corrected PD of 25\% of TDE2020mot \citep{TDE_2020mot_highest_pol}. Other explanations for optical emission of TDEs such as reprocessing of the inner disk remain plausible \citep{2014ApJ...783...23G, 10.1093/mnras/stac1727}.

Only about 1--10\% of detected TDEs exhibit a relativistic jet. The presence of such a jet can be inferred through synchrotron emission, which inevitably produces a polarized optical signal. Therefore, LAST-P can contribute to the rapid identification of transient sources as TDEs, which is crucial for enabling timely multi-wavelength follow-up observations. Interest in TDEs has also grown  within the VHE gamma-ray community \citep{10.1093/mnras/stw437}. While current-generation instruments have been actively searching for such events, none have been detected to date \citep[e.g., ][]{MAGIC_TDE, Lypova:2023qay}. TDEs have also been studied as potential sources of neutrinos via the acceleration of ultra-high-energy cosmic rays \citep{uhtde}.  

The wide-field monitoring capabilities of LAST-P will facilitate the discovery of nearby, bright TDEs (e.g., \citealt{2020ApJ...892L...1L}). Subsequent issued alerts would enable quick follow-up observations by the community. The high-cadence observing strategy of LAST-P will support simultaneous monitoring of the brightness evolution of multiple transients, contributing to estimates of the mass of the supermassive black hole of the host galaxy; population studies; and the classification of different TDE types. The Vera C. Rubin Observatory \citep{LSST} will survey the entire available southern sky every few nights, detecting up to $\sim$1000 TDEs per year \citep{LSST_tde}. The sub-day cadence of LAST-P will complement this survey, enabling time-resolved observations of the early evolution of the brightest TDEs. Assuming a TDE detection rate of approximately 50 events per year across 20,000\,deg$^2$ for sources brighter than g\,=\,19\,mag \citep[Fig.\,6 in][]{tde_rate}, we estimate that LAST-P, operating in high-cadence survey mode and covering about 31,000\,deg$^2$ of sky every 58 days (Table~\ref{tab:survey_summary}), could detect approximately 77 TDEs per year. The Transient Name Server \citep[TNS, ][]{TNS} lists 29 confirmed TDEs last year (Oct. 2024 -- Oct. 2025). In this context, we realistically expect to secure spectroscopic confirmation for 10--20 of our TDE candidates annually, assuming no major increase in spectroscopic resources. In deep mode, following up on confirmed TDEs, the PD uncertainty for a TDE with Gaia Bp-magnitude 18 would be as low as 0.8\%, with an MDP of 2.9\%.

\subsubsection{Fast Blue Optical Transients}

Fast Blue Optical Transients \citep[FBOTs, ][]{2018NatAs...2..307R,Margutti_2019}, are optical transients with a time above half-maximum of less than 12 days \citep{2014ApJ...794...23D}, usually fading within $30\,$days. Their radiated bolometric luminosities $L \approx 10^{43}$ erg s$^{-1}$ are detected in fast-cadence surveys. They are too fast and have weak hydrogen lines in their spectra, contrary to TDEs \citep{10.1093/mnrasl/slad145}.  FBOTs have mostly been associated with star-forming galaxies, suggesting collapsing massive stars as progenitors \citep{2014ApJ...794...23D}. A recent demographic study of FBOTs by \citet{2023ApJ...949..120H} found a diverse population. While some FBOTs are classified as hydrogen- or helium-rich core-collapse supernovae (Types II/IIb/Ib), interacting supernovae, wind shock breakout events \citep{Ofek_2010}, or events powered by low mass Ni$^{56}$ ejecta \citep{Ofek_2021}, some remains not well understood. Three of these events are particularly notable as their observed properties challenge standard supernova models. The observational characteristics of FBOTs are largely defined by a few archetypal events, beginning with AT2018cow (``the Cow''). These events are distinct from other luminous transients such as TDEs, primarily by their significantly faster evolution and bluer colors. FBOTs are distinguished by its rapid rise time ($t_{\rm rise} \approx 3.5$\,days), high peak luminosity ($L_{\rm peak} > 10^{44}$\,erg\,s$^{-1}$), and a luminous, persistent soft X-ray counterpart implying a long-lived central engine \citep{cow, cow_xray}. While sharing many of these characteristics, ZTF18abvkwla (``the Koala'') was exceptional for its radio emission, which was an order of magnitude more luminous than that of AT2018cow \citep{Ho_2020, 2023ApJ...949..120H}. The classification of these events as a distinct population was solidified with AT2020xnd (``the Camel''), the first unambiguous analogue to AT2018cow to be discovered and confirmed in real time. This event provided clear evidence for a powerful engine launching a relativistic outflow ($v > 0.5c$), as inferred from its exceptionally fast-evolving and luminous radio synchrotron emission \citep{10.1093/mnras/stab2785}. The extreme properties of these events suggest a powering mechanism beyond the radioactive decay typical of supernovae, pointing towards compact object engines or extraordinary circumstellar interactions.

Several physical interpretations have been proposed to explain their properties. One class of models involves the interaction of supernova ejecta with a dense circumstellar medium, leading to rapid deceleration and bright emission \citep{2014ApJ...780...18G,2018NatAs...2..307R}. Another related scenario involves mildly relativistic jets shocking their way out of the star, forming a surrounding "cocoon" \citep{2020ApJ...895L..23C,2022MNRAS.513.3810G}. Alternatively, the power source may be a compact central engine. Proposed engines include fall-back accretion onto the compact remnant \citep{2015MNRAS.451.2656K}, energy injection from a newborn millisecond magnetar \citep{2017ApJ...850...18H}, or accretion onto a neutron star or black hole \citep{2012MNRAS.420.2684C,2016MNRAS.461.1154M}.

For AT2018cow, observations revealed a polarization spike of $P=7\% \pm 1\%$ at $5.9\,$days post-burst, which later settled to a constant $P \approx 2\%$ \citep{10.1093/mnras/stad539}. This time-dependent polarization suggests an initially aspherical geometry, potentially explained by a shock breakout through an optically thick disk or a Compton-disk reflection model, the latter of which also accounts for an accompanying hard X-ray spike \citep{Margutti_2019}. A second FBOT with polarization measurement, AT2024wpp, showed a polarization of $P \leq 0.5\%$ between 6.1 and 14.4 days post-detection, consistent with the later, outflow-driven phase of AT2018cow \citep{10.1093/mnras/staf232}. However, constraining FBOT models is challenging due to their low detection rate of at most 0.1\% of the local core-collapse supernova rate \citep{2023ApJ...949..120H}. Large FoV surveys like LAST-P are therefore essential to systematically measure the polarization evolution of these events.

\subsection{Galactic Science}
\label{sect:galactic}

\subsubsection{Interstellar Medium}

In the optical to infrared range, polarization of starlight may arise from elongated dust grains that align with the local magnetic field. This phenomenon provides a powerful diagnostic for the geometry and strength of the Galactic magnetic field \citep[GMF, ][]{Andersson2015}, particularly when combined with precise astrometric distances from Gaia. 
Line-of-sight depolarization of starlight is a potential systematic effect, particularly for sightlines towards the Galactic plane or regions with complex magnetic topologies \citep{Panopoulou2022}.
Optical polarization measurements of individual stars are relatively unbiased, avoiding major limitations like the beam depolarization that affects radio observations of diffuse synchrotron emission.

Simultaneous measurements of the PD and PA for stars across a range of distances would enable us to create a three-dimensional (3D) tomographic map of dust-grain polarization. This, in turn, would reveal the structure of the GMF responsible for aligning those grains \citep{2024arXiv240910317M}. Regions where the PA varies rapidly with distance may indicate the presence of depolarizing layers; these can be masked to avoid contaminating the inferred magnetic field structure. Efforts in this direction have already been taken \citep{first_pol_tomography}. They have
led to the discovery of complex dust clouds
in a very limited region of the sky, which LAST-P
has the potential to build upon.


\subsubsection{Star-forming regions}

Young stellar objects (YSOs) such as Classical T~Tauri stars and Herbig Ae/Be stars often exhibit variable optical polarization due to dynamic changes in their circumstellar environments \citep{Manset2000, Vink2005}.
Gradual PA variations may indicate disk precession or evolving scattering angles, while stochastic PD changes can reveal clumpy accretion or inner disk shadowing effects. Optical polarimetry therefore constitutes an important probe of the geometry and evolution of these environments. 

For context,
Herbig Ae/Be stars have an average a PD of 3\% and a 1\,$\sigma$ spread of 3.4\% \citep{vink2015ttauri}. These stars have $R$-band AB magnitudes larger than magnitude 7.
Classical T~Tauri stars have an average PD of 1.6\% and 1\,$\sigma$ spread of 1.8\% \citep{vink2015ttauri}. Their $R$-band AB magnitudes generally also fall between 7 and 14 \citep{1994AJ....108.1906H}.
At magnitude 14, LAST-P can measure down to a PD of 0.7\% in 15\,x\,1 minute exposures, as part of our survey mode. This covers the vast majority of Herbig Ae/Be stars and of Classical~T Tauri stars. 

With its wide FoV and high-cadence survey mode, LAST-P would also be able to monitor entire star-forming regions. This capability enables the construction of polarization light curves for thousands of optically visible YSOs. In addition to studying known sources, LAST-P would identify new YSO candidates through polarization variability alone. This would compliment traditional infrared color-based selection methods.

\subsubsection{Galactic Novae}

Novae are explosive stellar events that provide key insights into stellar evolution \citep{BodeEvans2008}. Polarization observations help reveal the geometry and physical processes during the eruption. At peak brightness (typically magnitudes, 4--6), novae exhibit higher PD, within 1--5\%; this indicates asymmetric ejection of material \citep[e.g., ][]{Kawabata_2001}. As a nova evolves and the ejecta expand, the PD decreases to 0.2--1\% over a period of several months. This happens  due to the isotropization of the material, and to the cooling of the shell of the nova. In late phases, residual PD from the interstellar medium may still be detectable. Typically, a few months after the explosion, the $R$-band magnitude can reach values as faint as 10--14, depending on the distance, extinction, and on the intrinsic properties of the nova.

LAST-P, with its wide FoV, is well-suited for detecting novae across large regions of the sky. Despite their typically low polarization, novae are bright enough for LAST-P to measure PDs as small as 0.07\%  (magnitude 10 sources in a total of 15 minutes exposure split into smaller exposures to avoid saturation). While detection of polarization becomes more challenging in later stages, LAST-P can still measure PDs as low as 0.7\% for novae at magnitude 14.

\begin{figure}[]
\centering
\includegraphics[width=\linewidth]{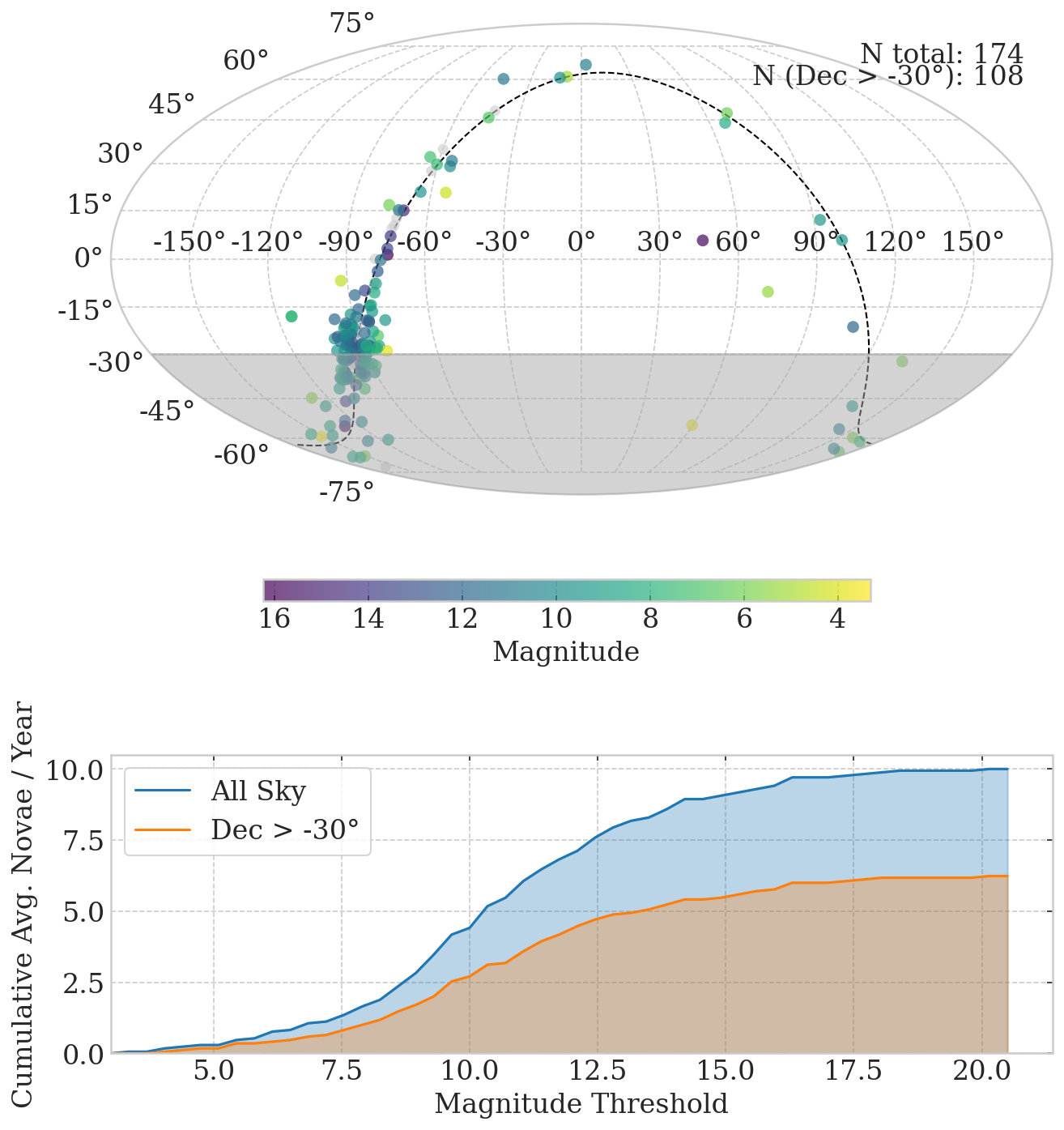}
\caption{Sky distribution (top) and cumulative discovery rates (bottom) for novae in the \textit{Koji} list of galactic Novae \cite{MukaiNovaeCatalog}. In the top panel, points with available V AB-magnitude are shown as part of the color scale. The Galactic Plane (dashed line) and the southern sky (Dec $<$\,-30$^\circ$; shaded gray area) are indicated. The bottom panel shows the average number of novae discovered per year (using data between 2008 and 2024) brighter than the magnitude threshold for the full sample (blue), as well as for a Dec $>$\,-30$^\circ$ subsample (orange).}
\label{fig:novae}
\end{figure}

The wide coverage and high temporal cadence mean that LAST-P is capable of detecting novae in their early phases, and of tracking their evolution over time. We expect to perform early detections of novae and to provide alerts to the community for follow-up observations. The estimated nova eruption rate in the Milky Way is $43.7^{+19.5}_{-8.7} \, \mathrm{yr}^{-1}$  \citep{novae_rate}. As discussed above, LAST-P would survey 75\% of the available sky every 18 days. We therefore expect to detect most novae that explode within the surveyed sky-region. For illustration, LAST-P would have had access to 62\% of the 174 novae detected between 2008 and 2024 by \cite{MukaiNovaeCatalog}, as shown in Fig.~\ref{fig:novae}.

\subsubsection{Galactic X-ray Binaries and microquasars}

X-ray binaries (XRBs) are systems where a compact object, such as a stellar-mass black hole or neutron star, accretes matter from a companion star, usually via an accretion disk. Many XRBs undergo transient outbursts. These are periods of dramatically increased luminosity across multiple wavelengths, presumably triggered by accretion instabilities. The latter fuel enhanced activity in the corona and are associated with relativistic jets \citep{Lipunov_2016}. High-energy X-ray emission is typically ascribed to the hot inner accretion flow or corona. Radio emission traces the large-scale jets. Optical and infrared (OIR) emission, on the other hand, reveals a more complex scenario. OIR light often contains contributions from the thermal emission of the accretion disk; the companion star itself; synchrotron emission from the base of the jet; and reprocessing of X-rays within the system. Hence, optical polarimetry provides a valuable probe of the geometry of accretion flows. It may be used to model the contribution of relativistic jets in X-ray binaries via scattering and synchrotron emission.

\begin{figure}
    \centering
    \includegraphics[width=\linewidth]{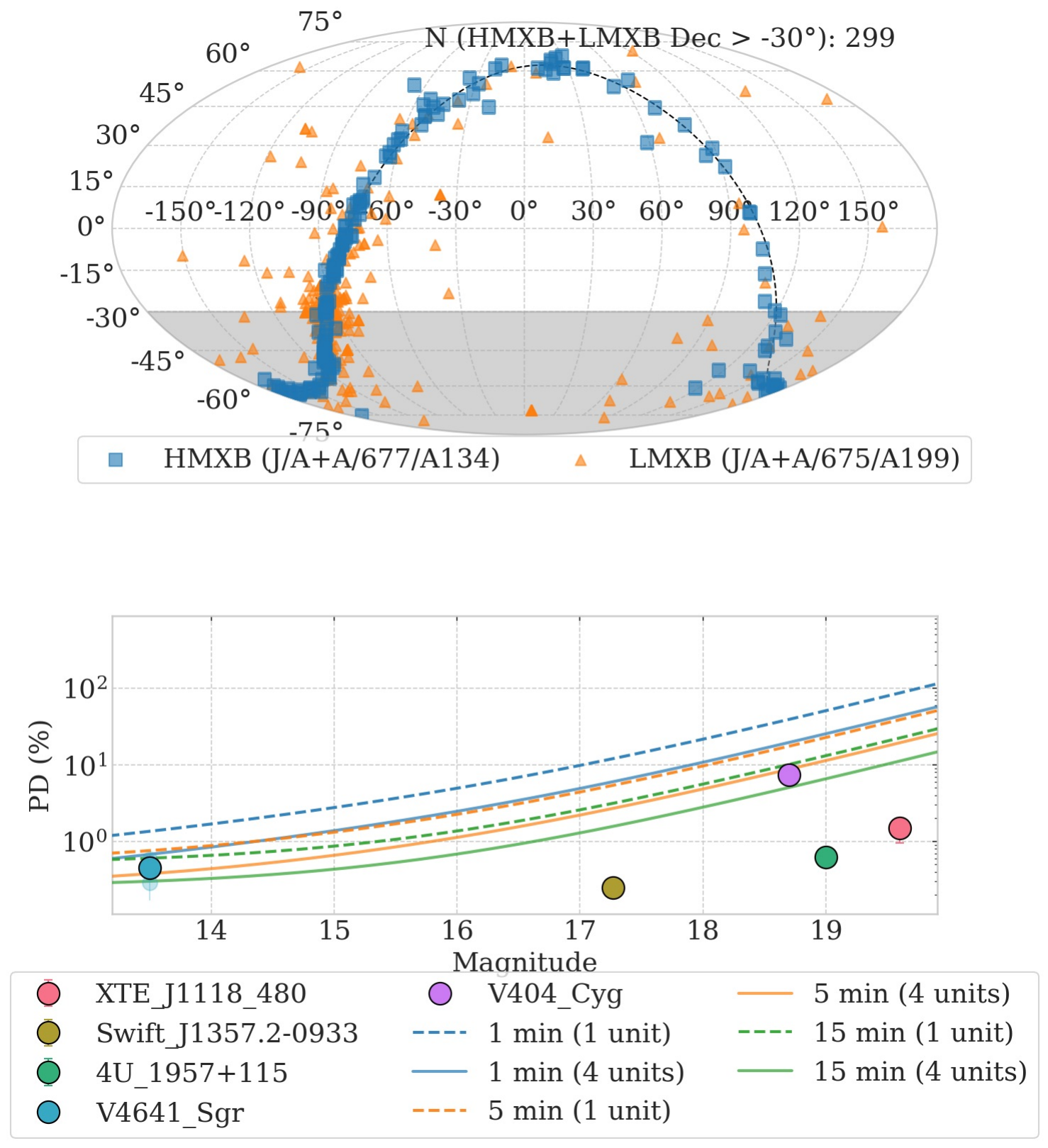}
    \caption{Top: Sky distribution of high-mass XRBs (HMXBs; blue squares) and low-mass XRBs \citep[LMXBs; orange triangles, ][]{xrb_catalog_1, xrb_catalog_2}. The Galactic Plane and Center are shown in black; the gray area indicates the region, Dec $<$ -30$^\circ$. Bottom: Polarization degree in the R band versus R magnitude for the sample of \citet{10.1093/mnras/stac1470}. Fainter points represent individual measurements; larger points indicate median or single-observation values per source. Dashed and solid lines respectively indicate the sensitivity (MDP) for our survey and deep modes of observation, given different exposure times.}
    \label{fig:xrb}
\end{figure}

Our benchmark for the expected magnitudes and polarization degrees of XRBs is \cite{10.1093/mnras/stac1470}. For such a sample, we find that only V4641 Sgr and V404 Cyg are detectable targets by a 15\,x\,1 minute exposure in survey mode. The other sources within this sample are predicted to be below the detection threshold of LAST-P. This is primarily due to their faintness or to their low intrinsic polarization. Nonetheless, the capabilities of LAST-P allow for a broad survey. Figure~\ref{fig:xrb} shows that 299 known XRBs are accessible to LAST-P, representing a large, unexplored population for optical polarimetry. Even in cases where our instrument would not significantly detect polarization, the measurements would constrain the potential contribution of non-stellar components to the overall spectra of XRBs.

Microquasars, a subset of X-ray binaries characterized by relativistic jets, are increasingly important targets for the gamma-ray astronomy community. Their jets can inflate extended radio lobes. They are assumed to create powerful shocks as they interact with the surrounding interstellar medium, accelerating particles to extreme energies. Such particle acceleration sites are responsible for the VHE gamma-ray emission detected from sources such as SS 433 by H.E.S.S. \citep{hess_ss433}. Other sources observed by LHAASO \citep{lhaaso} are also potentially of interest, e.g., making microquasars key science targets for the upcoming CTAO \citep{ctao_galactic}. Broad optical synchrotron lobes mirroring the gamma-ray structure are not visible in SS433. However, the jet-ISM interaction in the form of shock-excited filaments is seen in the optical regime within the W50 nebula. Measuring the corresponding PD and PA would allow us to map the magnetic field structures in the region.

\subsubsection{White Dwarfs}

White dwarfs (WDs) are dense remnants of low-to-intermediate mass stars. Due to their simple atmospheric composition and lack of strong intrinsic polarization mechanisms (in the non-magnetic majority), they are widely assumed to possess low (or zero) intrinsic polarization. 
Indeed, the majority of the PDs of WDs measured by RoboPol (Fig.~\ref{fig:wd_plots} bottom) are compatible with zero.
This makes WDs ideal standard stars for calibrating instrumental polarization. They can also be used to determine the foreground polarization induced by aligned dust grains in the ISM within our Galaxy.

As shown in Fig.~\ref{fig:wd_plots} (top), white dwarfs (WDs) are not strongly concentrated along the Galactic Plane due to their proximity to Earth. This wide spatial distribution is advantageous, providing a network of potential low-polarization calibrators across the entire sky \citep{10.1093/mnras/stw2335}.

A sub-type of WDs are magnetic WDs. These can exhibit polarization due to cyclotron emission or magneto-optical effects. Some WDs also possess dusty debris disks, where scattering can induce low levels of polarization \citep{wd}. Furthermore, WDs in close binary systems can exhibit polarization from scattering in accretion flows or outflows. The sensitivity of LAST-P reaches an MDP down to $\sim$0.87\% ($\sim$0.44\%) PD for stars brighter than magnitude $\approx 15$ in survey mode (deep mode) and 15\,x\,1 minute exposure. LAST-P therefore offers the capability to not only utilize numerous WDs for instrumental calibration across its FoV but also to search for subtle intrinsic polarization signals in WDs. This may potentially unveil weak magnetic fields, debris disks, or binary interactions. For illustration, the lower panel of Fig.~\ref{fig:wd_plots} shows the expected sensitivity of LAST-P relative to the measured polarization of the RoboPol sample \citep{10.1093/mnras/stw2335}.

\begin{figure}
    \centering
    \includegraphics[width=\linewidth]{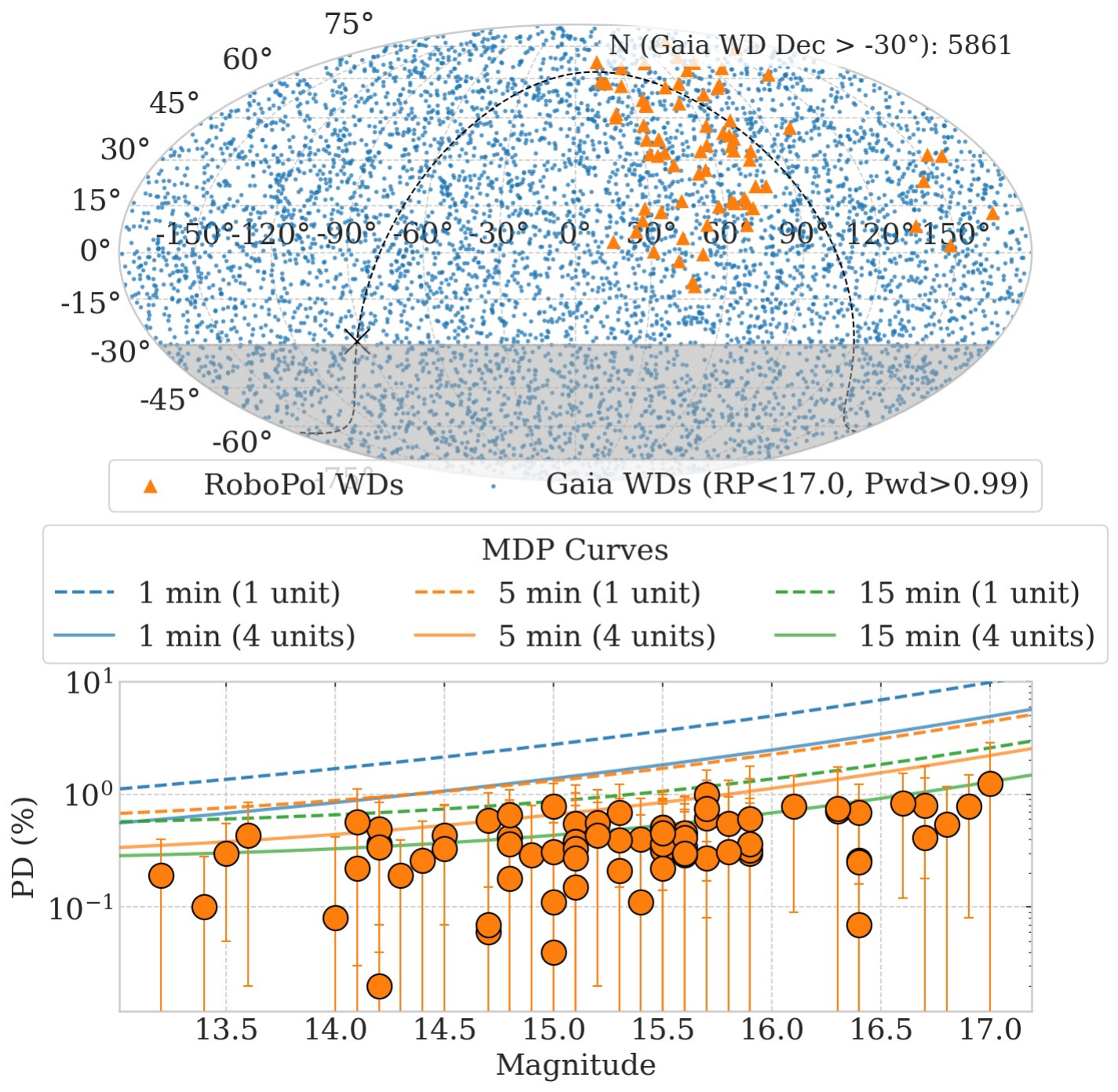}
    \caption{
    Top: Sky distribution of WDs, showing the RoboPol sample \citep[orange triangles, ][]{10.1093/mnras/stw2335}. High-probability (P$_{\rm WD} > 0.99$) WD candidates with RP magnitude smaller than 17 from the Gaia EDR3 catalog are indicated by blue markers \citep{10.1093/mnras/stab2672, vizier:J/MNRAS/508/3877}. The Galactic Plane and Center are respectively shown as a dark line and a shaded region. Bottom: PD \citep{10.1093/mnras/stw2335} versus Gaia DR3 RP magnitude. Lines represent the MDP curves for LAST-P given various exposure times, as indicated.
}
\label{fig:wd_plots}
\end{figure}

\section{Conclusions and summary}
\label{sect:conclusions}

In this paper, we presented LAST-P, the planned polarization node of the Large Survey Telescope. Our instrument is a cost-effective solution for building a novel wide-field telescope with polarimetric capabilities. LAST-P benefits from both the hardware and software developed for LAST \citep{Ofek_2023}, which is currently undergoing commissioning in Israel.

LAST-P will operate in two distinct modes. In survey mode, the 12 units, each composed of four telescopes, point at different sky regions, collectively covering 88.8\,deg$^2$. This mode enables full-sky coverage every 18--58 days depending on the cadence strategy. LAST-P can measure sources down to Bp-magnitude 20.9 in survey mode, depending on observation conditions. For magnitude 18 sources observed with a 15\,x\,1-minute exposure, the precision in the PD and PA reaches 1.5\% and 4.3$^\circ$, respectively, assuming a source with a PD of $\approx$ 10\% (Table~\ref{tab:pol_precision}). In deep mode, where four units simultaneously observe the same field, the improved SNR enhances sensitivity, achieving a PD precision of 0.75\%  (PA of 2.15$^\circ$) under similar conditions. Although initial tests have already been conducted, we plan a series of upgrades to the array, aimed at reducing overall instrumental polarization and minimizing systematic uncertainties in the polarization measurements.

LAST-P is designed to explore a diverse range of Galactic and extragalactic sources. Its high-cadence (sub-day) strategy will enable the monitoring of the largest sample of AGN in optical polarization to date. This will substantially increase the number of AGN with detected PA rotations. Such observations will provide critical insight into the connection between gamma-ray flares and PA variability. Polarimetric observations of GRBs from the earliest moments after the explosion will shed light on jet composition and structure. We also aim to investigate the outflows in supernovae (SNe) and the link between SNe and GRBs. Thanks to its wide FoV, LAST-P will contribute to the discovery and classification of rare transients, such as tidal disruption events (TDEs) and fast blue optical transients (FBOTs). Our detections will be coordinated with the broader multiwavelength and multi-messenger community in order to maximize their scientific potential.

LAST-P will also map the interstellar medium using polarized starlight. It will probe the polarization properties of variable stars such as Herbig Ae/Be and classical T Tauri stars, and will contribute to the largest catalog of low-polarization sources, focusing on white dwarfs. It will also study Galactic novae and X-ray binaries. This includes microquasars, where polarization data can reveal the geometry of ejected material and the formation mechanisms of relativistic jets.

A unifying theme across all these science cases is asymmetry; in outflows, in explosions, and as caused by jets. Polarized light offers a unique window into the physical geometry and emission mechanisms of such systems. LAST-P will thus make a significant contribution to our understanding of non-thermal astrophysical processes in the era of multi-messenger astronomy.

\begin{acknowledgments}
We thank the two anonymous referees for their comments, which helped us improve the focus and the quality of the manuscript.
We are grateful to Dr.~Ioannis Liodakis for his valuable attention and insightful suggestions on handling such a challenging dataset, provided during the HEPRO-9 conference in Rio de Janeiro. We thank A.~Krassilchtchikov for his insightful comments on the manuscript.
N.J-M. acknowledges support from the Alexander von
Humboldt Foundation.
S.G. is grateful for the support of the Koshland Family Foundation.
This work was supported by the Deutsche Forschungsgemeinschaft (DFG, German Research Foundation) through the Collaborative Research Center SFB1491 “Cosmic Interacting Matters- From Source to Signal” (grant no. 445052434)
E.O.O. is grateful for the support of grants from the Willner Family Leadership Institute, André Deloro Institute, Paul and Tina Gardner, The Norman E Alexander Family M Foundation ULTRASAT Data Center Fund, Israel Science Foundation, Israeli Ministry of Science, Minerva, BSF, BSF-transformative, NSF-BSF, Israel Council for Higher Education (VATAT), Sagol Weizmann-MIT, Yeda-Sela, and the Rosa and Emilio Segre Research Award. This research is supported by the Israeli Council for Higher Education (CHE) via the Weizmann Data Science Research Center, and by a research grant from the Estate of Harry Schutzman.
The authors acknowledge the use of Google Gemini 2.5 Pro for assistance with grammar and language editing in the preparation of this manuscript.
\end{acknowledgments}

\appendix

\section{Polarization and Uncertainty Equations}
\label{app:polarization}

In this appendix, we detail the analytical formulas used to derive the SNR, the polarization parameters, their corresponding uncertainties, and the detection limits from photometric intensity reference measurements \citep{LAST_pipeline}.

\subsection{SNR estimates for LAST-P}
\label{sec:snr_unified_model}
The polarimetric precision ($\sigma_{PD}$, $\text{MDP}$, $\sigma_{PA}$) is fundamentally governed by the SNR. We employ a $\text{SNR}$ estimate, which correctly combines all independent noise sources using the root sum square method.

The total $\text{SNR}$ for a source of magnitude $m$ in a co-added sequence is defined in terms of total collected electrons ($\text{e}^-$):
\begin{equation}
\text{SNR}(m) = \frac{N_*(m)}{\sqrt{N_*(m) + N_{\rm sky, new} + \sigma^2_{\rm fixed} + \sigma^2_{\text{sys}} (m)}}
\label{eq:snr_unified}
\end{equation}
where $N_*(m)$ are the source photon counts, $N_{\rm sky, new}$ are the background photon counts (reduced by filter throughput), $\sigma^2_{\rm fixed}$ is the variance contribution from non-photon noise (dark and readout noises), and $\sigma^2_{\text{sys}}$ is the variance contribution from the remaining systematic uncertainties.

\subsubsection{Derivation of Variables}
The parameters are calibrated assuming a dark-sky site ($21.5 \text{ mag arcsec}^{-2}$), an aperture of $2 \times \text{FWHM}$ ($A_{\rm aperture} \approx 91.6 \text{ arcsec}^2$ for the typical seeing of 2.7 arcsec), and a total exposure time of $T_{\rm total}=N_{\rm frames} \cdot t_{\rm exp}$. A filter throughput factor of $\eta_{\rm filter} = 0.1044$ is applied to all photon-dependent rates to account for the light absorption caused by the addition of a R-filter and a polarization filter (see Sect.~\ref{sec:transmittance}).

\begin{itemize}
    \item $\sigma^2_{\rm fixed}$: Total fixed noise variance. This variance represents the total contribution from detector effects across all co-added frames ($N_{\rm frames}=15$), remaining independent of the filter throughput. It combines the Dark Current ($R_{\rm dark} = 8 \times 10^{-3} \text{ e}^{-}/\text{pix}/\text{s}$) and the Read Noise \citep[$\sigma_{\rm read} = 2.7 \text{ e}^-$, Table 4 in ][]{LAST_pipeline}.

    First, the number of pixels in the aperture ($N_{\rm pix}$) is derived. The aperture radius is defined as $2 \times \text{FWHM}$ ($5.4 \text{ arcsec}$), which is converted to pixels using the system's $1.25 \text{ arcsec}/\text{pix}$ pixel scale, calculated as $N_{\rm pix} = \pi \cdot (5.4 \text{ arcsec} / 1.25 \text{ arcsec}/\text{pix})^2 \approx 58.6 \text{ pixels}$.

    The total fixed noise variance is then calculated:
    $$\sigma^2_{\rm fixed} = N_{\rm frames} \cdot N_{\rm pix} \cdot (\sigma_{\rm read}^2 + R_{\rm dark} \cdot t_{\rm exp}) \approx 6,836 \text{ e}^2$$

    \item $N_{\rm sky, new}$: Total Background Photon Counts. The surface brightness of the sky ($S_{\text{sky}} = 21.5 \text{ mag arcsec}^{-2}$) is converted to the total magnitude within the $A_{\rm aperture} \approx 91.6 \text{ arcsec}^2$ aperture ($m_{\rm sky, new}$):
    \begin{equation}
    \begin{split}
    m_{\rm sky, new} &= S_{\text{sky}} - 2.5 \log_{10}(A_{\rm aperture}) \\
    &\approx 21.5 - 2.5 \log_{10}(91.6) \approx 16.59 \text{ mag}
    \end{split}
    \end{equation}
    The initial unfiltered background counts ($N_{\rm sky}$) are derived by anchoring the system sensitivity to the observed single-frame $5\sigma$ limit of $m_{\lim}=19.6 \text{ mag}$ at $t_0=20 \text{ s}$ \citep{LAST_pipeline}. We account for +\,0.5 mag to account for the increase in sensitivity in regions with darker skies \citep{LAST_pipeline}, hence we start with $m_{\lim}=20.1 \text{ mag}$. The limiting rate $R_{\lim}$ is related to the unfiltered background rate $R_{\rm sky}$ by:
    $$R_{\lim} = R_{\rm sky} \cdot 10^{-0.4(m_{\lim} - m_{\rm sky, new})}$$
    The unfiltered background rate $R_{\rm sky}$ is found by solving the single-frame $\text{SNR}$ equation for $R_{\rm sky}$:
    \begin{equation}
    \label{eq:R_sky_solve}
    5 = \frac{R_{\lim} t_0}{\sqrt{(R_{\lim} + R_{\rm sky}) t_0 + \sigma^2_{\rm fixed, 1}}}
    \end{equation}
    where $R_{\lim}$ is substituted using the magnitude relationship. The variables used in the equation solution are: $\sigma^2_{\rm fixed, 1}$ is the single-frame fixed variance, calculated as $\sigma^2_{\rm fixed, 1} = N_{\rm pix} \cdot (\sigma_{\rm read}^2 + R_{\rm dark} \cdot t_0) \approx 436 \text{ e}^2$ using $t_0 = 20 \text{ s}$, $N_{\rm pix} = 58.6$, and the detector parameters. This solution yields $R_{\rm sky} \approx 847.7$ and the total unfiltered background counts $N_{\rm sky} = R_{\rm sky} \cdot T_{\rm total} \approx 762,970 \text{ e}^-$. This value is then scaled by $\eta_{\rm filter}$:
    $$N_{\rm sky, new} = N_{\rm sky} \cdot \eta_{\rm filter} \approx 79,654 \text{ e}^-$$
    The combined background noise ''floor'' used in the denominator is $N_{\rm sky, new} + \sigma^2_{\rm fixed} \approx 86,490 \text{ e}^2$.

    \item $Z_{\rm R, new}$: The zero point rate is a constant that defines the source flux for $m = 0$. Hence, the source flux for a source magnitude $m$ is $R_*(m) = Z_{\rm R, new} \cdot 10^{-0.4m}$. $Z_{\rm R, new}$ is calculated from the unfiltered rate ($R_{\rm sky}$) and the aperture magnitude ($m_{\rm sky, new} \approx 16.59 \text{ mag}$), and then scaled by $\eta_{\rm filter}$:
    $$Z_{\rm R, new} = R_{\rm sky} \cdot 10^{0.4 \cdot m_{\rm sky, new}} \cdot \eta_{\rm filter} \approx 3.85 \times 10^{8} \text{ e}^{-}/\text{s}$$

    Therefore, the  the source flux for a source magnitude $m$ is $R_*(m) \approx 3.85 \times 10^{8} \cdot 10^{-0.4m} \text{ e}^{-}/\text{s}$ and the source photon counts for a total exposure of $T_\mathrm{total} = 900 s$ is $N_* (m) \approx 3.465 \times 10^{11 - 0.4m} \text{ e}^{-}$.

    \item $\sigma^2_{\text{sys}}$: The systematic uncertainties. For this initial analysis, the systematic error variance ($\sigma^2_{\text{sys}}$) is estimated using the LAST pipeline's reported bright-end relative photometric precision. \citet{Ofek_2023} states a relative precision of $3 \text{ millimag}$ in a single $20 \text{ s}$ image, but to represent the stability inherent in the co-added sequence, we adopt a standard deviation of $\sigma_{m, \text{sys}} = 1.5 \text{ millimag}$ for the magnitude measurements. The variance converts this magnitude uncertainty into the electron count domain ($\text{e}^2$) using the expression:
$$\sigma_\text{sys}^2 (m) = \left[ N_*(m) \cdot \frac{\ln(10)}{2.5} \cdot (1.5 \times 10^{-3}) \right]^2.$$

This $\text{SNR}$ model assumes controlled systematic uncertainties; additional errors from the full photometric/polarimetric pipeline stability, as well as system instabilities, will be quantified in future work following instrument commissioning.
    
\end{itemize}

Substituting the numerical values for the $15 \times 60 \text{ s}$ co-add:
\begin{equation}
\label{eq:master}
\text{SNR}(m) = \frac{3.465 \times 10^{11 - 0.4m}}{\sqrt{3.465 \times 10^{11 - 0.4m}  + 2.29 \times 10^{17 - 0.8m} + 86490}}
\end{equation}

\subsection{Exact Data Reduction Formulas}

We calculate the normalized Stokes parameters, $q$ and $u$, from intensity measurements taken at four polarization filter angles ($I_{0}, I_{45}, I_{90}, I_{135}$):
\begin{equation}
    q = \frac{I_{0} - I_{90}}{I_{0} + I_{90}}
    \label{eq:stokes_q}
\end{equation}
\begin{equation}
    u = \frac{I_{45} - I_{135}}{I_{45} + I_{135}}
    \label{eq:stokes_u}
\end{equation}

The PD and PA are derived as:
\begin{equation}
    PD \, (\%) = 100 \times \sqrt{q^2 + u^2}
    \label{eq:pd}
\end{equation}
\begin{equation}
    PA = \frac{1}{2} \arctan\left(\frac{u}{q}\right)
    \label{eq:pa}
\end{equation}
where the PA is typically adjusted to the range $[0^\circ, 180^\circ)$.

To determine the measurement errors during data reduction, we propagate the photometric flux uncertainties ($\sigma_{I_{\alpha}}$) into the Stokes parameters:
\begin{equation}
    \sigma_q = \frac{2}{(I_0+I_{90})^2} \sqrt{ (I_{90}\sigma_{I_0})^2 + (I_{0}\sigma_{I_{90}})^2 }
    \label{eq:err_q}
\end{equation}
\begin{equation}
    \sigma_u = \frac{2}{(I_{45}+I_{135})^2} \sqrt{ (I_{135}\sigma_{I_{45}})^2 + (I_{45}\sigma_{I_{135}})^2 }
    \label{eq:err_u}
\end{equation}

The uncertainty in PD ($\sigma_{PD}$) is calculated from $\sigma_q$ and $\sigma_u$:
\begin{equation}
    \sigma_{PD} = \frac{\sqrt{(q\sigma_q)^2 + (u\sigma_u)^2}}{\sqrt{q^2+u^2}}
    \label{eq:err_pd}
\end{equation}

Finally, the uncertainty in the polarization angle ($\sigma_{PA}$), in radians, is:
\begin{equation}
    \sigma_{PA} = \frac{1}{2(q^2+u^2)}\sqrt{(u\sigma_q)^2 + (q\sigma_u)^2}
    \label{eq:err_pa}
\end{equation}

\subsection{Minimum degree of polarization (MDP)}
\label{app:mdp_def}

To quantify the sensitivity limit of the instrument, we define the Minimum Detectable Polarization (MDP). Following \citet{10.1117/12.857357} and \citet{2012SPIE.8443E..4NE}, the MDP is given by:

\begin{equation}
\label{eq:mdp_def}
\mathrm{MDP}= \sigma\,1.43\,\frac{\sigma_{\text{tot}}}{S_{\text{tot}}}\times100 ,
\end{equation}

\noindent where:
\begin{itemize}
  \item \( \sigma \) is the chosen significance level (e.g.\ \(\sigma=2.576\) for a 99\% c.l. limit);
  \item \( S_{\text{tot}} = \displaystyle\sum_{i=1}^{4}\mathrm{counts}_{\text{cam}i}\) is the total source signal from the four cameras;
  \item \( \sigma_{\text{tot}} = \sqrt{\displaystyle\sum_{i=1}^{4}\sigma_i^{2}}\) is the combined uncertainty (including background) added in quadrature.
\end{itemize}
This formulation includes a factor of 1.43 to account for the Rician bias inherent in measuring the positive-definite polarization degree at low SNR ratios.

\subsection{Sensitivity Estimation approximations}
\label{app:sensitivity_approx}

To estimate the theoretical capabilities of LAST-P (Section \ref{sec:instr_photometry}), we assume that the source has low intrinsic polarization ($I_0 \approx I_{90} \approx I_{tot}/2$). Under these conditions, the error propagation formulas above simplify to functions of the photometric SNR.

Defining the total signal $S_{\text{tot}}$ and total uncertainty $\sigma_{\text{tot}}$ over the aperture, the ratio $\sigma_{\text{tot}} / S_{\text{tot}}$ simplifies to the inverse of the global SNR:
\begin{equation}
    \frac{\sigma_{\text{tot}}}{S_{\text{tot}}} \approx \frac{1}{\mathrm{SNR}}
\end{equation}

Substituting this approximation into Eq. \ref{eq:mdp_def}, the MDP simplifies to a direct function of the SNR and the required confidence level ($\sigma$):
\begin{equation}
    \mathrm{MDP} \approx 1.43 \times \sigma \times \frac{1}{\mathrm{SNR}} \times 100
    \label{eq:mdp_approx}
\end{equation}
For a 99\% confidence level ($\sigma = 2.576$), this constant factor is 3.684.

Similarly, for the PD uncertainty, defining the total intensity $I_{tot} = I_0 + I_{90}$ and assuming uniform noise $\sigma_{I} \approx \sqrt{I_{tot}}$, Eq. \ref{eq:err_q} approximates to:
\begin{equation}
    \sigma_q \approx \sigma_u \approx \frac{\sigma_{\text{tot}}}{S_{\text{tot}}} \approx \frac{1}{SNR}
\end{equation}

Substituting these approximations into Eq. \ref{eq:err_pd}, the absolute uncertainty in the PD simplifies to:
\begin{equation}
    \sigma_{PD} \approx \frac{\sqrt{q^2(1/SNR)^2 + u^2(1/SNR)^2}}{\sqrt{q^2+u^2}} = \frac{1}{SNR}
    \label{eq:approx_pd}
\end{equation}

For the PA, substituting into Eq. \ref{eq:err_pa} yields:
\begin{equation}
    \sigma_{PA} \approx \frac{1}{2(q^2+u^2)} \frac{1}{SNR} \sqrt{u^2 + q^2} = \frac{1}{2 \cdot PD \cdot SNR}
    \label{eq:approx_pa}
\end{equation}

This demonstrates that while $\sigma_{PD}$ depends only on the signal strength, $\sigma_{PA}$ is inversely proportional to the source's intrinsic polarization. These approximated forms (Eqs. \ref{eq:mdp_approx}, \ref{eq:approx_pd} and \ref{eq:approx_pa}) are used to generate the theoretical precision estimates in Table \ref{tab:pol_precision}.

\subsection{Additional sensitivity estimates}
\label{app:complete_table}

\begin{table*}
    \centering
    \caption{Estimated sensitivity of LAST-P grouped by exposure time for several magnitudes.}
    \label{tab:complete_table_updated}
    
    \begin{tabular}{c c c c c c}
        \toprule
        \textbf{mag} & \textbf{Exposure} & \textbf{SNR} & \textbf{PD uncert (\%)} & \textbf{PA uncert (deg)} & \textbf{MDP (\%)} \\ 
        \midrule
        \multirow{3}{*}{8} & 30 s & 700 & 0.14 & 0.41 & 0.53 \\
        & 60 s & 711 & 0.14 & 0.40 & 0.52 \\
        & 900 s & 723 & 0.14 & 0.40 & 0.51 \\
        \midrule
        \multirow{3}{*}{12} & 30 s & 366 & 0.27 & 0.8 & 1.0 \\
        & 60 s & 462 & 0.22 & 0.6 & 0.8 \\
        & 900 s & 691 & 0.14 & 0.4 & 0.5 \\
        \midrule
        \multirow{3}{*}{15} & 30 s & 95 & 1.0 & 3.0 & 3.9 \\
        & 60 s & 133 & 0.7 & 2.1 & 2.8 \\
        & 900 s & 426 & 0.2 & 0.7 & 0.9 \\
        \midrule
        \multirow{3}{*}{16} & 30 s & 53 & 1.9 & 5.4 & 6.9 \\
        & 60 s & 75 & 1.3 & 3.8 & 4.9 \\
        & 900 s & 270 & 0.4 & 1.1 & 1.4 \\
        \midrule
        \multirow{3}{*}{17} & 30 s & 27 & 3.8 & 10.8 & 14 \\
        & 60 s & 38 & 2.7 & 7.6 & 9.8 \\
        & 900 s & 143 & 0.7 & 2.0 & 2.6 \\
        \midrule
        \multirow{3}{*}{18} & 30 s & 12.1 & 8.2 & 24 & 30 \\
        & 60 s & 17.1 & 5.8 & 17 & 21 \\
        & 900 s & 66.1 & 1.5 & 4.3 & 5.6 \\
        \midrule
        \multirow{3}{*}{19} & 30 s & 5.1 & 19 & 56 & 72 \\
        & 60 s & 7.3 & 13.7 & 40 & 51 \\
        & 900 s & 28.2 & 3.5 & 10 & 13 \\
        \midrule
        \multirow{3}{*}{20} & 30 s & 2.1 & - & - & - \\
        & 60 s & 3.0 & - & - & - \\
        & 900 s & 11.5 & 9 & 25 & 32 \\
        \midrule
        \bottomrule
    \end{tabular}
    \par\medskip
    \small
    \emph{Note:} $\sigma_{PA}$ values are based on the small-angle approximation and on a source with PD of 10\%. MDP is calculated at the 99\% confidence level from Eq.~\ref{eq:mdp_approx}. Values are not shown for SNR$ < $5. The limiting magnitudes for the total exposures of 30, 60 and 900\,s are 19.0, 19.4, and 20.9, respectively.
\end{table*}

\begin{figure}
    \centering
    \includegraphics[width=1\linewidth]{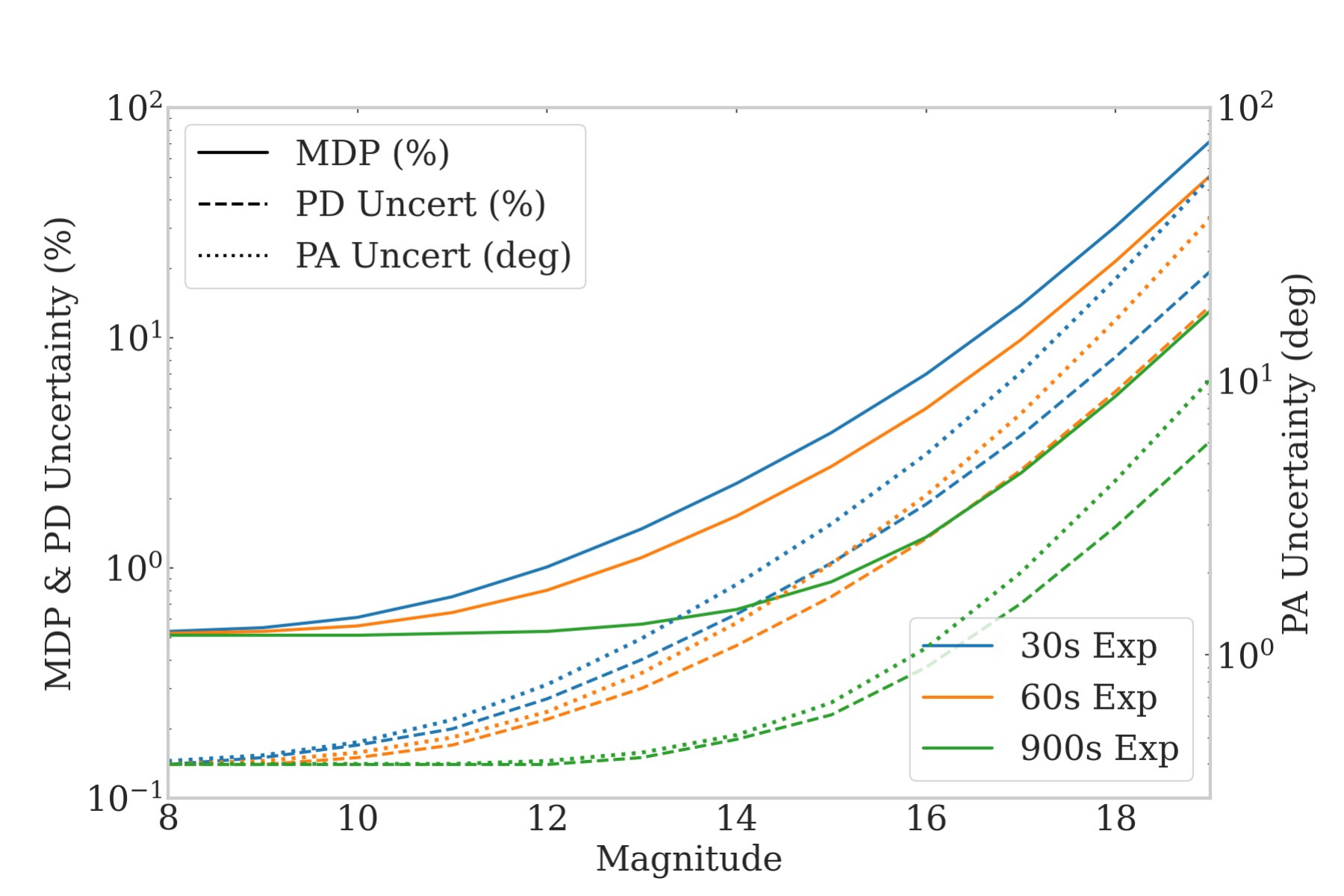}
\caption{Polarimetric sensitivity as a function of source magnitude for three different exposure times ($30$, $60$, and $900$\,s). The left $y$-axis displays the MDP (solid lines) and the uncertainty in PD ($\sigma_{PD}$, dashed lines) in percentage. The secondary $y$-axis (right) shows the uncertainty in PA ($\sigma_{PA}$, dotted lines) in degrees.}    \label{fig:sensitivity}
\end{figure}

We utilize Eq.~\ref{eq:snr_unified} to estimate the SNR for some magnitudes for different exposures and Eqs.~\ref{eq:pd_main_text}, ~\ref{eq:pa_main_text}, ~\ref{eq:mdp_main_text} for estimating the PD, PA, and MDP, respectively. These estimates are shown in Table~\ref{tab:complete_table_updated} and for better visualization in Fig.~\ref{fig:sensitivity}. The systematic uncertainties, governed by the adopted photometric stability limit of $\sigma_{m, \text{sys}} = 1.5 \text{ millimag}$, introduce a sensitivity ''floor'' that restricts the best achievable precision to $\text{SNR}_{\text{max} \approx 720}$, resulting in polarization limits of $\sigma_{PD} \approx 0.15\%$, $\sigma_{PA} \approx 0.40^\circ$ (for $PD=10\%$), and an $\text{MDP} \approx 0.51\%$. 

\bibliography{sample701}{}
\bibliographystyle{aasjournalv7}



\end{document}